\documentclass[%
 reprint,
 superscriptaddress,
 amsmath,amssymb,
prl,
]{revtex4-2}

\usepackage{graphicx}%
\usepackage{dcolumn}%
\usepackage{bm}%
\usepackage{physics}
\usepackage[dvipsnames]{xcolor}
\usepackage[
    colorlinks,
    citecolor=RoyalBlue,
    linkcolor=RoyalBlue,
    urlcolor=RoyalBlue,
]{hyperref}
\usepackage{multirow}
\usepackage{mathtools}
\usepackage{verbatim}  %
\usepackage[normalem]{ulem}

\newcommand{\red}[1]{\textcolor{Maroon}{#1}}
\newcommand{\blue}[1]{\textcolor{RoyalBlue}{#1}}

\newcommand{\affilITMO}{School of Physics and Engineering, ITMO University, St.~Petersburg 197101, Russia}
\newcommand{\affilANU}{Nonlinear Physics Center, Research School of Physics, Australia National University, Canberra ACT 2601, Australia}
\newcommand{\affilRIGA}{Riga Technical University, Institute of Telecommunications, Riga 1048, Latvia}
\newcommand{\affilMIPT}{Center for Photonics and 2D Materials, Moscow Institute of Physics and Technology, Dolgoprudny 141700, Russia}

\begin{document}

\title{
Nonlinearity-induced optical torque
}

\author{Ivan Toftul}
\email{toftul.ivan@gmail.com}
\affiliation{\affilANU}
\affiliation{\affilITMO}

\author{Gleb Fedorovich}
\affiliation{\affilITMO}

\author{Denis Kislov}
\affiliation{\affilITMO}
\affiliation{\affilRIGA}
\affiliation{\affilMIPT}

\author{Kristina Frizyuk}
\affiliation{\affilITMO}

\author{Kirill Koshelev}
\affiliation{\affilANU}

\author{Yuri Kivshar}
\affiliation{\affilANU}

\author{Mihail Petrov}
\affiliation{\affilITMO}

\date{\today}

\begin{abstract}
Optically-induced mechanical torque leading to the rotation of small objects requires the presence of absorption or breaking cylindrical symmetry of a scatterer. A spherical non-absorbing particle cannot rotate due to the conservation of the angular momentum of light upon scattering. Here, we suggest a novel physical mechanism for the angular momentum transfer to non-absorbing particles via nonlinear light scattering. The breaking of symmetry occurs at the microscopic level manifested in {\it nonlinear negative optical torque} due to the excitation of resonant states at the harmonic frequency with higher projection of angular momentum. The proposed physical mechanism can be verified with resonant dielectric nanostructures, and we suggest some specific realizations. 

\end{abstract}

\maketitle

\paragraph*{Introduction.} The rotation and spinning of  micro- and nanoscale objects is one of the central goals of optical manipulation since the discovery of optical tweezers~\cite{Ashkin1970Jan,Ashkin1971Oct,Ashkin1974Jun,Ashkin1975Mar,Ashkin1986May,Chu1985Jul,Gordon1980May,Ashkin1978Mar}, utilized in controlling biological systems~\cite{Zhang2008Apr,Dholakia2020Sep,Fazal2011Jun}, atoms~\cite{Kaufman2021Dec,Kim2019Apr}, and nanoscale objects~\cite{Marago2013Nov,Shi2022Sep,Kostina2019Mar,Tkachenko2020Jan,Toftul2020Jan}.  
The transfer of  angular momentum from light to matter results in a mechanical torque acting on a scatterer~\cite{jackson1998ClassicalElectrodynamics,novotny2010PrinciplesNanoOptics,L.Allen2014Apr,Ye2017May}, being proportional to a difference between the  angular momenta absorbed and re-scattered by the object. nonzero mechanical torque can appear due to the lack of rotational symmetry~\cite{Brasselet2009Jun,Friese1998Jul,Simpson2007Oct,Trojek2012Jul} or the presence of absorption~\cite{Marston1984Nov,Canaguier-Durand2013Sep}.
The direction and sign of the mechanical torque is defined by the imbalance condition, and  it can be opposite to the projection of the incident angular moment of light leading to {\it negative optical torque} (NOT)~\cite{Chen2014Sep}. The appearance of linear NOT has recently been studied both theoretically~\cite{Chen2014Sep,Mitri2016Oct,Nieto-Vesperinas2015Jul} and experimentally~\cite{Han2018Nov,Sule2017Nov,Diniz2019Mar,Hakobyan2014Aug}. 

\begin{figure}[t]
\centering
\includegraphics[width=0.7\linewidth]{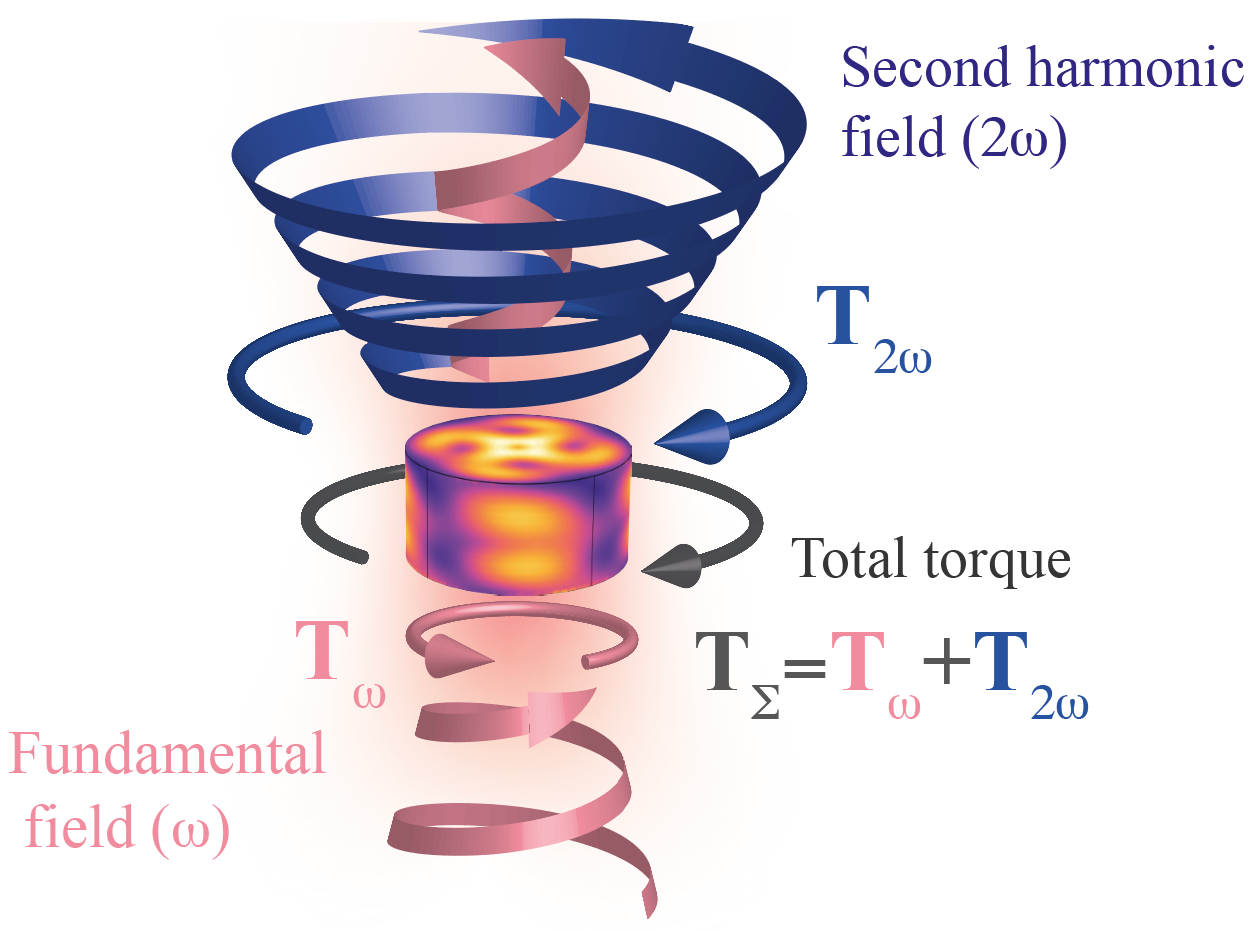}
\caption{General concept. Circular polarized light at the frequency $\omega$ is launched onto a cylindrical dielectric particle and generates second-harmonic fields at the frequency $2\omega$ that might have different angular momentum due to a crystalline lattice structure, producing a nonlinearity-induced optical torque enhanced by the Mie resonances.}
\label{fig:general_fig}
\end{figure}

Rapid  development of all-dielectric nanophotonics~\cite{Zograf2021Sep,Baranov2017Jul,Kivshar2018Mar,Krasnok2015May} brings novel opportunities for optical manipulation.  In contrast to nanoplasmonics, dielectric materials have lower Ohmic losses~\cite{Decker2016Sep}, which are required for realizing optical rotation of cylindrically symmetric  structures~\cite{Marston1984Nov,Canaguier-Durand2013Sep}. 
However, dielectric structures offer unique opportunities for observing  nonlinear optical processes such as second harmonic generation (SHG) or third-harmonic generation (THG) due to large values of bulk nonlinear susceptibilities. 
It is also possible to observe  experimentally SHG in trapped particles~\cite{Malmqvist1995Jun,Sato1994Jul}.
Recently, the dramatic enhancement of the SHG efficiency for resonant all-dielectric nanostructures was reported~\cite{Frizyuk2019Feb,Koshelev2021Jan,Carletti2019Sep,Carletti2018Jul,Bonacina2020Jun}. 
Here, we suggest utilizing the SHG for a transfer of angular momenta of light to scatterers via nonlinear optical process. 
The generated \textit{second harmonic (SH) field also may  carry the angular momenta} and, thus, provides a contribution to the mechanical torque. We predict that the angular momentum imbalance between the fields  at the fundamental and SH frequencies can lead to an optical torque even for non-absorbing particles with cylindrical symmetry (Fig.~\ref{fig:general_fig}), and  
its  sign can change from positive to \textit{negative} with respect to the incident field  angular momentum.

\paragraph*{Nonlinearity-induced optical torque.} 
We start with considering   circularly polarized  plane wave   with frequency $\omega$ scattered  on a dielectric particle possessing the azimuthal symmetry (see Fig.~\ref{fig:general_fig}). 
The plane wave is incident along the axis of the symmetry and carries the momentum of light of $m_{\text{inc}}\hbar$ per photon. 
Due to the symmetry of the problem the optical torque  $\vb T^{(\omega)}$ acting on the particle at the fundamental frequency is exactly proportional to the absorption cross section~\cite{Marston1984Nov,Canaguier-Durand2013Sep} and, in terms of canonical spin angular momenta density, one can write
$\vb{T}^{(\omega)} = {c}/{n_0} \cdot \sigma_{\text{abs}} \vb{S}^{(\omega)}$, where $\vb{S}^{(\omega)} = m_{\text{inc}}/{(2\omega)} \cdot \varepsilon \varepsilon_0 [E_0^{(\omega)}]^2\vb e_z$ is the 
canonical spin angular momenta density~\cite{Bliokh2017Aug} with azimuthal number $m_{\text{inc}}=\pm 1$ for right(left) circular polarization and $n_0=\sqrt{\varepsilon \mu}$ is the refractive index of the host media; 
$\sigma_{\text{abs}}$ is the total absorption cross section. 
For a nonlinear process in dielectric particles, the energy loss at the fundamental harmonic (FH) frequency occurs due to the harmonic generation, so $\sigma_{\text{abs}} \to \sigma_{\text{SHG}}$.
With the consideration of SHG being a dominant nonlinear process~\cite{Ok2006Jul}, one should also account for the angular momenta carried out by the SH field (Fig.~\ref{fig:general_fig}). 
Hence, there are two components of the optical nonlinear torque
\begin{equation}
    \vb{T} = \vb{T}^{(\omega)} + \vb{T}^{(2\omega)},
\end{equation}
where $\vb{T}^{(\omega)}$ and $\vb{T}^{(2\omega)}$ are the torques generated by the field at FH and SH. The interference terms with nonzero frequencies are averaged to zero~\cite{SM}.

For the particles possessing an azimuthal symmetry with negligible Ohmic losses 
excited at frequencies
away from two and three photon absorption regions~\cite{Hurlbut2007Mar}, torque on the FH is defined by the amount of energy spent on the SHG process. 
By decomposing SH field into the series of vector spherical harmonics (VSH) it is possible to express SH generation cross section $\sigma_{\text{SHG}}$ in terms of the radial electromagnetic energy density in magnetic and electric multipoles in the far field $W^{\text E}_{mj}$ and $W^{\text M}_{mj}$~\cite{SM,Frizyuk2019Feb}, so torque at the FH is
\begin{equation}
    T_z^{(\omega)} = m_{\text{inc}} T_0 \frac{\sigma_{\text{SHG}}}{\sigma_{\text{geom}}}
    =  \frac{m_{\text{inc}} T_0}{\sigma_{\text{geom}} [k(2\omega)]^2} \sum_{jm} \left( W^{\text E}_{mj}+W^{\text M}_{mj} \right),
    \label{eq:torque1w_sum}
\end{equation}
where $k(2\omega) = n_0 2\omega /c$, $\sigma_{\text{geom}}$ is the geometric cross section, $mj$ are the projection and the total angular momentum numbers, and $T_0= 0.5 \varepsilon \varepsilon_0 [E_0^{(\omega)}]^2 \sigma_{\text{geom}} / k(\omega)$, is the maximal torque which can transferred to a plate of area $\sigma_{\text{geom}}$ once all the momentum of the incident field is absorbed.
Coefficients $W^{\text E}_{mj}$ and $W^{\text M}_{mj}$ depend on the overlap integral between the nonlinear polarization $\vb P^{(2\omega)} = \varepsilon_0 \hat{\chi}^{(2)} \vb E^{(\omega)} \vb E^{(\omega)}$ and field of the SH mode in the volume of the particle~\cite{SM}. 
Thus $W^{\text E}_{mj}$ and $W^{\text M}_{mj}$ contains all the information about nonlinear  response.

\begin{figure}
\centering
\includegraphics[width=1.0\linewidth]{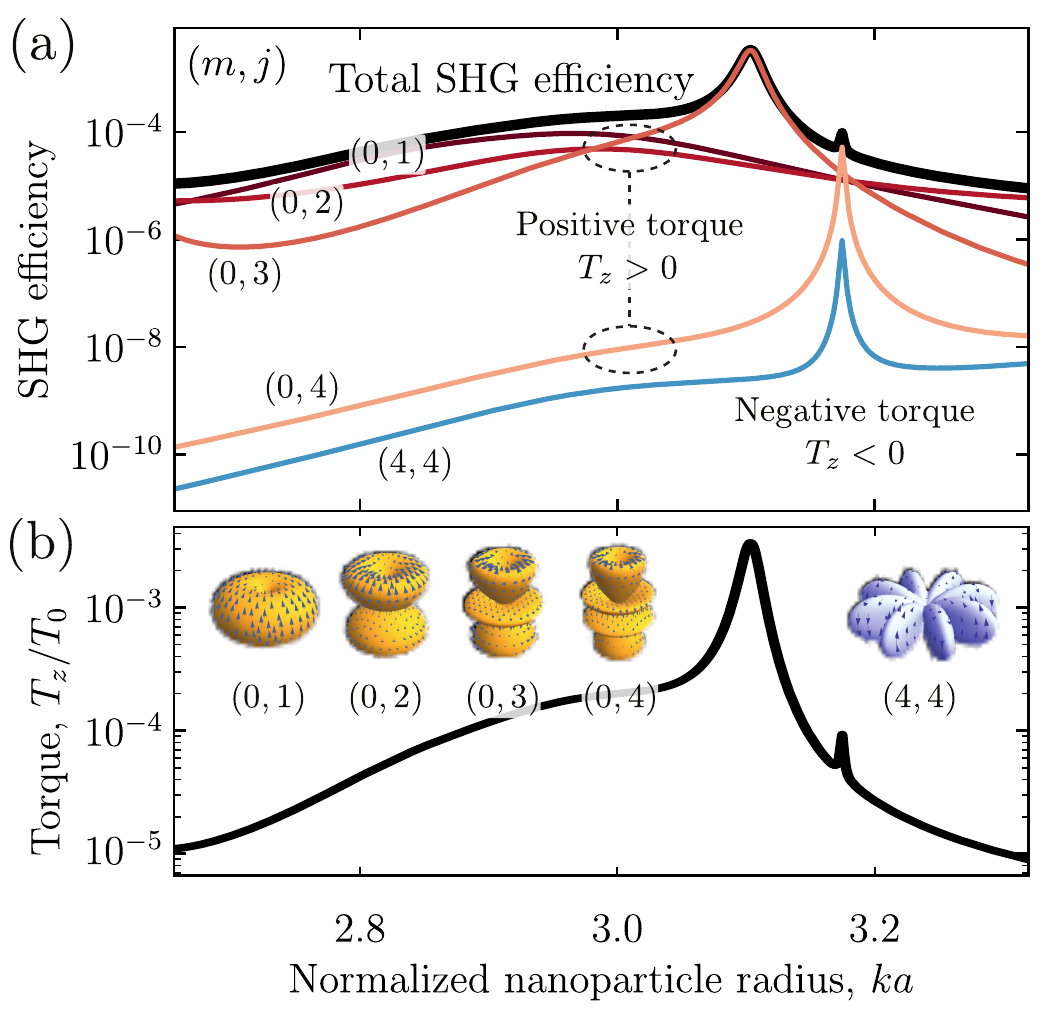}
\caption{Exact solution for a dielectric sphere. (a) SHG efficiency ($\sigma_{\text{SHG}}/\sigma_{\text{geom}}$) of a spherical GaAs nanoparticle with refractive indices $n^{(\omega)}_{\text{p}} = 3.28$, $n^{(2\omega)}_{\text{p}} = 3.56$ as a function of its radius. 
The colored lines show the contribution of different multipolar channels labeled as $(m,j)$.
The pump wavelength is 1550 nm. The multipoles with $m=0$ and $m=4$ provide contribution to positive and negative optical torques, respectively. (b) Total optical torque acting on the nanoparticle due to SHG.} 
\label{fig:sphere_torque}
\end{figure}

The torque on the SH frequency can be derived by the calculating the change of the total angular momenta flux tensor $\vb{\hat{\mathcal{M}}}^{(2\omega)}$ on the SH as $\vb{T}^{(2\omega)} = \oint_\Sigma  \vb{\hat{\mathcal{M}}}^{(2\omega)} \cdot \vb{n} \dd S$~\cite{novotny2010PrinciplesNanoOptics,Ye2017May,griffiths2005introduction, jackson1998ClassicalElectrodynamics,Mun2020Sep,bliokh2014ExtraordinaryMomentumSpin}. 
Surface integration is performed over arbitrary closed surface $\Sigma$ which contains the scatterer, and $\vb{n}$ is the outer normal to that surface. This integral can be taken once the fields are decomposed into the VSH series~\cite{Li2014Dec,Li2012Jul,Pesce2020Dec,Borghese2006Oct}
and the total torque can be written in a compact and elegant way which underpins the physics behind nonlinearity-induced optical torque~\cite{SM}:
\begin{eqnarray} \label{eq:tot_torque}
    T_z &=& \red{T_z^{(\omega)}} + \blue{T_z^{(2\omega)}}
    \\ \nonumber
    &=& \frac{1}{2} T_0 \frac{1}{\sigma_{\text{geom}} [k(2\omega)]^2} \sum_{jm} (\red{2m_{\text{inc}}}  - \blue{m}) \left[ W^{\text E}_{mj}+W^{\text M}_{mj} \right].
\end{eqnarray}
Eq.~\eqref{eq:tot_torque} is the central result of this work. 
Generation of SH photon in the particular VSH state  results in \textit{(i)} adding a torque corresponding to the spins of two photons absorbed at the FH and \textit{(ii)} adding a recoil torque from  SH photons emitted with the total angular momentum projection $m$.
This nonlinear optomechanical effect  has not been discussed in the literature and is proposed for the first time, to the best of our knowledge.       

\paragraph*{Selection rules.} 
For the in-depth analysis of Eq.~\eqref{eq:tot_torque} we use the symmetry analysis and multipolar decomposition~\cite{Tsimokha2022Apr,Sadrieva2019Sep,Frizyuk2019Aug,Frizyuk2021May,Frizyuk2019Feb}. 
The imbalance between the angular momentum projections in Eq.~\eqref{eq:tot_torque} immediately shows that there can appear a nonzero torque induced by nonlinear optical generation process results. 
Its sign with respect to $m_{\text{inc}}$ strongly depends on the exact multipolar content of SH field. 
Most of these components are zero due to the strict selection rules on $m$ during SHG~\cite{Frizyuk2019Feb,Frizyuk2019Aug,Finazzi2007Sep,Makarov2017May,Smirnova2018Jan} imposed by the symmetry of the particle and the crystalline lattice, which can be explicitly seen from the overlapping integral in $W_{mj}^{\text{E,H}}$~\cite{SM}.
In order to illustrate the mechanism of NOT appearance, we consider individual crystalline structures made of GaAs which is a common optical material possessing strong second-order nonlinear optical response. 
Its zinc-blende lattice structure with  T$_\text d$ symmetry provides a single independent component of the nonlinear tensor  $\chi^{(2)}_{xyz}$\cite{Smith1958-MacroscopicSymmetry,boyd2020nonlinear, Miller1964-OPTICALSECONDHARMON,Franken1963-OpticalHarmonicsand} once the lattice is oriented such that $[001] \parallel \vb{\hat{e}}_z, [100] \parallel \vb{\hat{e}}_x$. 
\begin{figure}[t]
\centering
\includegraphics[width=1.00\linewidth]{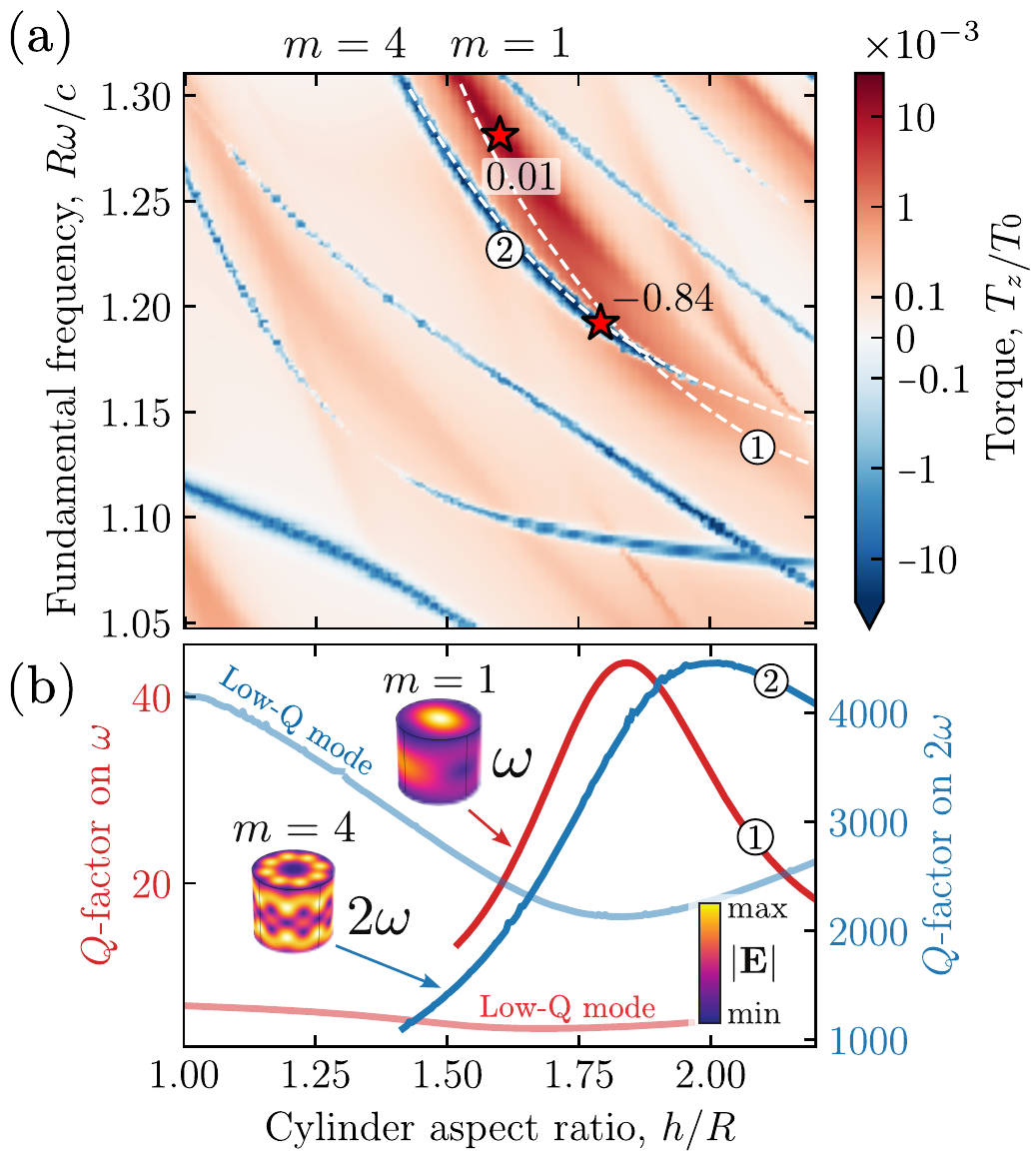}
\caption{
(a) Total spinning torque on a cylinder with a fixed radius $R = 250$~nm made of GaAs with refractive index $n_\text{cyl}=3.5$ placed in air as a function of dimensionless fundamental frequency $R \omega/c$ and aspect ratio $h/R$, where $h$ is the cylinder height. 
Two red stars shows maximal positive and maximal negative values of the torque. 
Notably, the maximal negative torque coincide with the intersection of the two eigenmodes on FH and SH (white dashed lines), which gives double resonant condition for the nonlinearity-induced optical torque.
(b) The $Q$-factors of these eigenmodes are shown, which have $m=4$ on SH and $m=1$ on FH.
Light blue and light red lines show the $Q$-factors of the leaky modes near high-$Q$ modes.
Electric field amplitude is $E_0^{(\omega)} = 8.68 \cdot 10^7~\text{V/m}$.
}
\label{fig:Cyl_torque}
\end{figure}

The symmetry of GaAs lattice along with the axial symmetry of the nanoparticle dictates that only $m=0, \pm4$ for incident RCP (LCP) wave are allowed in the SH field, i.e. $2m_{\text{inc}}$ from the incident field and $\pm 2$ from the $\hat{\chi}^{(2)}_{\text{GaAs}}$ tensor (see Supplemental Material for $\hat{\chi}^{(2)}_{\text{GaAs}}$ explicitly written in cylindrical coordinates~\cite{SM} and also Refs.~\cite{Frizyuk2021-NonlinearCircularDi,Frizyuk2019Feb}). Now from Eq.~\eqref{eq:tot_torque} one can see that for the RCP incident field, which has azimutal number $m_{\text{inc}}=1$, harmonics with $m=0$  provide a positive contribution to the total torque since $2m_{\text{inc}}-m=2$ ($ T_z^{(\omega)}>0$, $T_z^{(2\omega)}=0$), while with $m=4$ provide a negative contribution since $2m_{\text{inc}}-m=-2$ ($T_z^{(2\omega)}=-2 T_z^{(\omega)}$) and the recoil torque at SH  overcomes the torque at FH. 

We next apply these selection rules to SHG in a spherical GaAs particle (including dispersion relations).  
The particle radius was chosen in the range from 200 nm to 250 nm at the excitation wavelength of 1550 nm. 
For chosen parameters the excitation is resonant with magnetic dipole mode at the fundamental frequency.  
The SH field has dominant $j=1\dots4$ multipolar terms shown in Fig.~\ref{fig:sphere_torque} (b) inset, and their contribution in the overall SHG power are shown in Fig.~\ref{fig:sphere_torque} (a), where SHG efficiency $\sigma_{\text{SHG}}/\sigma_{\text{geom}}$ is plotted. 
The magnetic and electric multipole counterparts are not specified in the plot.
As discussed above only $m=0$ and $m=4$ components are  present in the SH field. 
One can see that the major contribution to the SHG signal is governed by the harmonics with $m=0$, while the hexadecapolar harmonic $j=4, m=4$ providing a negative torque contribution is very weak. Thus, the SHG in GaAs spherical particles results in {\it positive} optical torque (see Fig.~\ref{fig:sphere_torque} (b)) reproducing the SH efficiency spectra according to Eq.~\eqref{eq:tot_torque}.

\paragraph*{Negative optical torque enabled by high-$Q$ modes.}  The multipole content of SH field can be modified and controlled  by designing the resonator shape~\cite{Gladyshev2020,Sadrieva2019,Liu2020}. In the view of this work, we aim at enhancing the contribution of $m=4$ modes in the SH field, which can enable appearance  of {\it negative} optical torque. 
That can be achieved by 
lifting degeneracy between modes with different $m$,
utilizing Mie modes with the high total angular momentum $j$ and quasi-bound states in the continuum (qBIC) with high-$Q$ factors and ability to enhance light-matter interaction in the nanoscale structures~\cite{Koshelev2022Jul,Rybin2017Dec,Koshelev2020Jan,Bogdanov2019Jan}.
The qBIC states can be easily observed in cylindrical particles by variation of height to radius ratio  preserving the axial symmetry of the system. 
The Friedrich-Wintgen  mechanism~\cite{FriedrichH.1985} allows interaction of different modes with the same $m$ via the radiation continuum resulting in the formation of high-$Q$ qBIC modes along with low-$Q$ modes.
Fig.~\ref{fig:Cyl_torque} (b) shows the  $Q$-factor of the resonant modes tuned at the FH with $m=1$ and at SH with $m=4$, which provides a double resonance condition.  The dominating contribution of qBIC modes in the SH spectrum immediately leads to the appearance of negative optical torque (see Fig.~\ref{fig:Cyl_torque}~(a)) close to $m=4$ modes excitation as soon as the recoil contribution of SHG with $m=4$ prevails over the torque due to generation of mode with $m=0$. Moreover, the resonant character of the effect provides switching of the optical torque from positive to negative in a very narrow  range of parameters. 

\begin{figure}[t]
\centering
\includegraphics[width=0.94\linewidth]{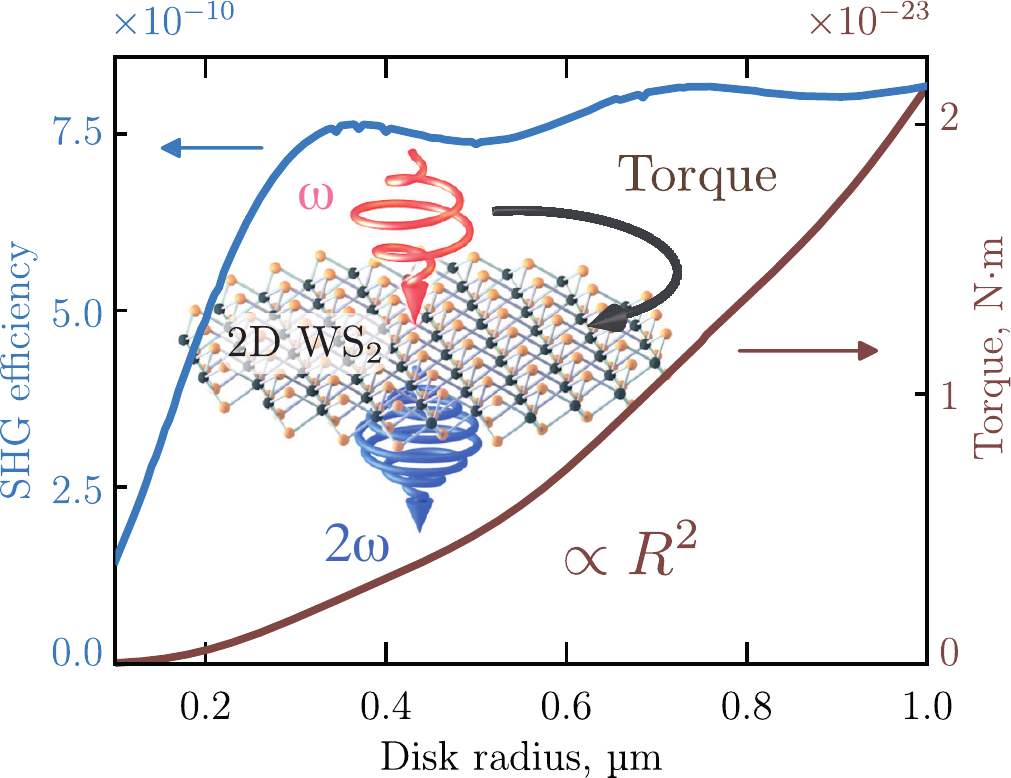}
\caption{SHG generation and induced optical torque for WS$_2$ flakes. Shown are SHG efficiency $\sigma_{\text{SHG}}/\sigma_{\text{geom}}$ as a function of the flake size (blue) and induced nonlinear torque (brown). Intensity of the incident field is $E_0^{(\omega)} = 8.68 \cdot 10^7~\text{V/m}$, and wavelength is 1550~nm. Inset: proposed geometry for experimental realization.}
\label{fig:2D_torque}
\end{figure}

Excitation of high-$Q$ resonant states leads to the drastic increase of the second harmonic efficiency and optical torque in accordance to   Eq.~\eqref{eq:torque1w_sum}. The  estimation based on the coupled mode theory~\cite{Koshelev2020Jan} gives (see SM~\cite{SM}):
\begin{equation}
    \frac{\sigma_{\text{SHG}}}{\sigma_{\text{geom}}} \simeq 10^{-8} Q_1^2 Q_2 \frac{{I}^{(\omega)}_0}{1[\text{GW/cm}^2]},
    \label{eq:eff_est}
\end{equation}
where ${I}^{(\omega)}_0 \approx 1.3~\text{GW}/\text{cm}^2$ is the intensity of the pump field, the $Q_{1,2}$ are the qBICs $Q$-factors on the FH and SH, correspondingly (see Fig.~\ref{fig:Cyl_torque}). In the steady-state regime, the optical torque leads to the rotation of the object at a constant frequency limited by the viscous friction. For a dielectric cylinder in water estimations  give us  $\Omega_{\text{cyl}} \sim 10^{5}~{\text{rad}}/{\text{s}}$, 
for the double resonance condition with $Q_1 \approx 50$ and $Q_2 \approx 1000$.  

\paragraph*{Rotation of TMDC flakes.} 
The rotation motion of cylindrical structures is unstable for $h/R \gg 1$, losing its top-like stability and preventing experimental observation of the proposed effect. The critical AS can be found by setting the equal transverse and longitudinal moments of inertia, which gives $h/R = \sqrt{3}$. (see SM~\cite{SM}). In this view, the rotation of atomically thin non-absorptive disks  made of two-dimensional (2D) material is deprived of such limitation. The transition metal dichalogenides (TMDC)  are known as an efficient platform for flat nonlinear optics~\cite{McDonnell2016Oct}. The recent experimental successes in TMDC stable floating on top of liquids~\cite{Zhao2020Nov,Jin2019Jul} inspires us for suggesting the potential geometry of the experiment shown in Fig.~\ref{fig:2D_torque}. The circle  structure cut of 2D TMDC flake floating over liquid and illuminated by the laser and resulting in SHG. The D$_{3 \text h}$  symmetry of the crystalline lattice provides the following nonzero components of nonlinear tensor $\chi^{(2)}_{xxx}=-\chi^{(2)}_{xyy}=-\chi^{(2)}_{yyx}=-\chi^{(2)}_{yxy}=\chi^{(2)}_{2D}$~\cite{boyd2020nonlinear}, which provide that SH modes with $m=\mp 1$ are dominantly generated under  right(left) $m_{\text{inc}}=\pm 1$  circular excitation~\cite{boyd2020nonlinear}. Then the difference between the  angular momenta of the generated SH field and the incident field is always negative and the torque is directed along the with incident angular momenta (positive optical torque) and equals to $T_z = \pm 2{c}/{n_0} \cdot \sigma_{\text{SHG}}$. In Fig.~\ref{fig:2D_torque} the  dependence of the SHG efficiency on the radius of the  TMDC structure is shown. 
The pump wavelength is $1550$~nm with the pump power flux of $2$~GW/cm$^2$ which also lies below the two-photon absorption threshold. 
The nonlinear tensor coefficient is $\chi^{(2)}_{2D}=50$ pm/V~\cite{You2019Jan}, while the refractive index at the FH and SH frequency is $2.75$ and $3.12$ respectively~\cite{Ermolaev2019Dec} which corresponds to WS$_2$ material. 
From Fig.~\ref{fig:2D_torque}~(b) one can see that the SHG efficiency tends to a constant value with the increase of the structure size corresponding to the efficiency of an infinite sheet. 
The torque related to the generation of the SH increases quadratically with the structure size.
The estimation of the rotation frequency gives $\Omega_{\text{2D}} \sim 0.1~{\text{rad}}/{\text{s}}$~\cite{SM}.

\paragraph*{Discussions.} 
The proposed mechanism of nonlinear optical torque occurs to the breaking cylindrical symmetry at the microscopic level, i.e. accounting for the crystal lattice symmetry. 
Indeed, in the SHG process the rule of the momentum projection conservation is satisfied~\cite{Nikitina2022-Whendoesnonlinearc} with correction to additional momentum provided by the susceptibility tensor written in the cylindrical coordinates~\cite{SM}. The proposed mechanism of nonlinearity-induced torque can be observed via other nonlinear processes such as the THG or higher-harmonic generation.  For example, for silicon nanostructures  (O$_{\text h}$ lattice symmetry) the SHG process is suppressed while THG is quite strong. The $\hat \chi ^{(3)}$ tensor provides the additional momentum $m=\pm4$ in the third harmonic field in full analogy to the SHG. We also should stress that the THG in isotropic materials will not induce optical torque as the $\hat \chi ^{(3)}$ tensor has specific form~\cite{boyd2020nonlinear} which does not give additional angular momenta projection to the field.

Finally, the generation of a nonlinear torque requires quite strong excitation intensities up to $1$ GW/cm$^2$ . For such intensities, multiphoton absorption can contribute into overall losses leading to a parasitic optical torque not related to the generation of nonlinear fields.  Thus, the proposed mechanism in can become possible once two-photon energy is below the band gap  $2\hbar\omega<E_g$, and two-photon absorption is suppressed which is for room temperature $E_g^{\text{GaAs}} \simeq 1.4~\text{eV}$ corresponding to $872~\text{nm}$.
 
\paragraph*{Conclusion.} 
We have presented the general theory of nonlinearity-induced optical torque,
originating from the  angular momentum transfer from the photonic field to a nanostructure via harmonics generation.  We have demonstrated that a nonzero optical torque can appear for the case of a non-absorptive dielectric structures with a rotational symmetry,
and the resulting angular frequency can be as high as 100 kHz. Additionally, the stable rotation of circular TMDC flakes of single-layer WS$_2$ under the circularly polarized light excitation is also possible. We believe that our work paves a way towards novel intriguing phenomena driven by nonlinearity-induced optomechanical manipulation.

\begin{acknowledgements}
We thank R.~Quidant and A.~Solntsev for useful comments, and K.~Dholakia for references.  
The work is partially supported by the Program Priority 2030.  
D.K. acknowledges 
the Latvian Council of Science (project No. lzp-2021/1-0048).  
I.T. and Y.K. acknowledge a support fro the Australian Research Council (the grant DP210101292), the International Technology Center Indo-Pacific (ITC IPAC) and Army Research Office (contract No. FA520921P0034). 
\end{acknowledgements}

\bibliography{torque_SHG_bib}

\begin{thebibliography}{110}%
\makeatletter
\providecommand \@ifxundefined [1]{%
 \@ifx{#1\undefined}
}%
\providecommand \@ifnum [1]{%
 \ifnum #1\expandafter \@firstoftwo
 \else \expandafter \@secondoftwo
 \fi
}%
\providecommand \@ifx [1]{%
 \ifx #1\expandafter \@firstoftwo
 \else \expandafter \@secondoftwo
 \fi
}%
\providecommand \natexlab [1]{#1}%
\providecommand \enquote  [1]{``#1''}%
\providecommand \bibnamefont  [1]{#1}%
\providecommand \bibfnamefont [1]{#1}%
\providecommand \citenamefont [1]{#1}%
\providecommand \href@noop [0]{\@secondoftwo}%
\providecommand \href [0]{\begingroup \@sanitize@url \@href}%
\providecommand \@href[1]{\@@startlink{#1}\@@href}%
\providecommand \@@href[1]{\endgroup#1\@@endlink}%
\providecommand \@sanitize@url [0]{\catcode `\\12\catcode `\$12\catcode
  `\&12\catcode `\#12\catcode `\^12\catcode `\_12\catcode `\%12\relax}%
\providecommand \@@startlink[1]{}%
\providecommand \@@endlink[0]{}%
\providecommand \url  [0]{\begingroup\@sanitize@url \@url }%
\providecommand \@url [1]{\endgroup\@href {#1}{\urlprefix }}%
\providecommand \urlprefix  [0]{URL }%
\providecommand \Eprint [0]{\href }%
\providecommand \doibase [0]{https://doi.org/}%
\providecommand \selectlanguage [0]{\@gobble}%
\providecommand \bibinfo  [0]{\@secondoftwo}%
\providecommand \bibfield  [0]{\@secondoftwo}%
\providecommand \translation [1]{[#1]}%
\providecommand \BibitemOpen [0]{}%
\providecommand \bibitemStop [0]{}%
\providecommand \bibitemNoStop [0]{.\EOS\space}%
\providecommand \EOS [0]{\spacefactor3000\relax}%
\providecommand \BibitemShut  [1]{\csname bibitem#1\endcsname}%
\let\auto@bib@innerbib\@empty
\bibitem [{\citenamefont {Ashkin}(1970)}]{Ashkin1970Jan}%
  \BibitemOpen
  \bibfield  {author} {\bibinfo {author} {\bibfnamefont {A.}~\bibnamefont
  {Ashkin}},\ }\bibfield  {title} {\bibinfo {title} {{Acceleration and Trapping
  of Particles by Radiation Pressure}},\ }\href
  {https://doi.org/10.1103/PhysRevLett.24.156} {\bibfield  {journal} {\bibinfo
  {journal} {Phys. Rev. Lett.}\ }\textbf {\bibinfo {volume} {24}},\ \bibinfo
  {pages} {156} (\bibinfo {year} {1970})}\BibitemShut {NoStop}%
\bibitem [{\citenamefont {Ashkin}\ and\ \citenamefont
  {Dziedzic}(1971)}]{Ashkin1971Oct}%
  \BibitemOpen
  \bibfield  {author} {\bibinfo {author} {\bibfnamefont {A.}~\bibnamefont
  {Ashkin}}\ and\ \bibinfo {author} {\bibfnamefont {J.~M.}\ \bibnamefont
  {Dziedzic}},\ }\bibfield  {title} {\bibinfo {title} {{Optical Levitation by
  Radiation Pressure}},\ }\href {https://doi.org/10.1063/1.1653919} {\bibfield
  {journal} {\bibinfo  {journal} {Appl. Phys. Lett.}\ }\textbf {\bibinfo
  {volume} {19}},\ \bibinfo {pages} {283} (\bibinfo {year} {1971})}\BibitemShut
  {NoStop}%
\bibitem [{\citenamefont {Ashkin}\ and\ \citenamefont
  {Dziedzic}(1974)}]{Ashkin1974Jun}%
  \BibitemOpen
  \bibfield  {author} {\bibinfo {author} {\bibfnamefont {A.}~\bibnamefont
  {Ashkin}}\ and\ \bibinfo {author} {\bibfnamefont {J.~M.}\ \bibnamefont
  {Dziedzic}},\ }\bibfield  {title} {\bibinfo {title} {{Stability of optical
  levitation by radiation pressure}},\ }\href
  {https://doi.org/10.1063/1.1655064} {\bibfield  {journal} {\bibinfo
  {journal} {Appl. Phys. Lett.}\ }\textbf {\bibinfo {volume} {24}},\ \bibinfo
  {pages} {586} (\bibinfo {year} {1974})}\BibitemShut {NoStop}%
\bibitem [{\citenamefont {Ashkin}\ and\ \citenamefont
  {Dziedzic}(1975)}]{Ashkin1975Mar}%
  \BibitemOpen
  \bibfield  {author} {\bibinfo {author} {\bibfnamefont {A.}~\bibnamefont
  {Ashkin}}\ and\ \bibinfo {author} {\bibfnamefont {J.~M.}\ \bibnamefont
  {Dziedzic}},\ }\bibfield  {title} {\bibinfo {title} {{Optical Levitation of
  Liquid Drops by Radiation Pressure}},\ }\href
  {https://doi.org/10.1126/science.187.4181.1073} {\bibfield  {journal}
  {\bibinfo  {journal} {Science}\ }\textbf {\bibinfo {volume} {187}},\ \bibinfo
  {pages} {1073} (\bibinfo {year} {1975})}\BibitemShut {NoStop}%
\bibitem [{\citenamefont {Ashkin}\ \emph {et~al.}(1986)\citenamefont {Ashkin},
  \citenamefont {Dziedzic}, \citenamefont {Bjorkholm},\ and\ \citenamefont
  {Chu}}]{Ashkin1986May}%
  \BibitemOpen
  \bibfield  {author} {\bibinfo {author} {\bibfnamefont {A.}~\bibnamefont
  {Ashkin}}, \bibinfo {author} {\bibfnamefont {J.~M.}\ \bibnamefont
  {Dziedzic}}, \bibinfo {author} {\bibfnamefont {J.~E.}\ \bibnamefont
  {Bjorkholm}},\ and\ \bibinfo {author} {\bibfnamefont {S.}~\bibnamefont
  {Chu}},\ }\bibfield  {title} {\bibinfo {title} {{Observation of a single-beam
  gradient force optical trap for dielectric particles}},\ }\href
  {https://doi.org/10.1364/OL.11.000288} {\bibfield  {journal} {\bibinfo
  {journal} {Opt. Lett.}\ }\textbf {\bibinfo {volume} {11}},\ \bibinfo {pages}
  {288} (\bibinfo {year} {1986})}\BibitemShut {NoStop}%
\bibitem [{\citenamefont {Chu}\ \emph {et~al.}(1985)\citenamefont {Chu},
  \citenamefont {Hollberg}, \citenamefont {Bjorkholm}, \citenamefont {Cable},\
  and\ \citenamefont {Ashkin}}]{Chu1985Jul}%
  \BibitemOpen
  \bibfield  {author} {\bibinfo {author} {\bibfnamefont {S.}~\bibnamefont
  {Chu}}, \bibinfo {author} {\bibfnamefont {L.}~\bibnamefont {Hollberg}},
  \bibinfo {author} {\bibfnamefont {J.~E.}\ \bibnamefont {Bjorkholm}}, \bibinfo
  {author} {\bibfnamefont {A.}~\bibnamefont {Cable}},\ and\ \bibinfo {author}
  {\bibfnamefont {A.}~\bibnamefont {Ashkin}},\ }\bibfield  {title} {\bibinfo
  {title} {{Three-dimensional viscous confinement and cooling of atoms by
  resonance radiation pressure}},\ }\href
  {https://doi.org/10.1103/PhysRevLett.55.48} {\bibfield  {journal} {\bibinfo
  {journal} {Phys. Rev. Lett.}\ }\textbf {\bibinfo {volume} {55}},\ \bibinfo
  {pages} {48} (\bibinfo {year} {1985})}\BibitemShut {NoStop}%
\bibitem [{\citenamefont {Gordon}\ and\ \citenamefont
  {Ashkin}(1980)}]{Gordon1980May}%
  \BibitemOpen
  \bibfield  {author} {\bibinfo {author} {\bibfnamefont {J.~P.}\ \bibnamefont
  {Gordon}}\ and\ \bibinfo {author} {\bibfnamefont {A.}~\bibnamefont
  {Ashkin}},\ }\bibfield  {title} {\bibinfo {title} {{Motion of atoms in a
  radiation trap}},\ }\href {https://doi.org/10.1103/PhysRevA.21.1606}
  {\bibfield  {journal} {\bibinfo  {journal} {Phys. Rev. A}\ }\textbf {\bibinfo
  {volume} {21}},\ \bibinfo {pages} {1606} (\bibinfo {year}
  {1980})}\BibitemShut {NoStop}%
\bibitem [{\citenamefont {Ashkin}(1978)}]{Ashkin1978Mar}%
  \BibitemOpen
  \bibfield  {author} {\bibinfo {author} {\bibfnamefont {A.}~\bibnamefont
  {Ashkin}},\ }\bibfield  {title} {\bibinfo {title} {{Trapping of Atoms by
  Resonance Radiation Pressure}},\ }\href
  {https://doi.org/10.1103/PhysRevLett.40.729} {\bibfield  {journal} {\bibinfo
  {journal} {Phys. Rev. Lett.}\ }\textbf {\bibinfo {volume} {40}},\ \bibinfo
  {pages} {729} (\bibinfo {year} {1978})}\BibitemShut {NoStop}%
\bibitem [{\citenamefont {Zhang}\ and\ \citenamefont
  {Liu}(2008)}]{Zhang2008Apr}%
  \BibitemOpen
  \bibfield  {author} {\bibinfo {author} {\bibfnamefont {H.}~\bibnamefont
  {Zhang}}\ and\ \bibinfo {author} {\bibfnamefont {K.-K.}\ \bibnamefont
  {Liu}},\ }\bibfield  {title} {\bibinfo {title} {{Optical tweezers for single
  cells}},\ }\href
  {https://royalsocietypublishing.org/doi/full/10.1098/rsif.2008.0052}
  {\bibfield  {journal} {\bibinfo  {journal} {J. R. Soc. Interface}\ }\textbf
  {\bibinfo {volume} {5}},\ \bibinfo {pages} {671} (\bibinfo {year}
  {2008})}\BibitemShut {NoStop}%
\bibitem [{\citenamefont {Dholakia}\ \emph {et~al.}(2020)\citenamefont
  {Dholakia}, \citenamefont {Drinkwater},\ and\ \citenamefont
  {Ritsch-Marte}}]{Dholakia2020Sep}%
  \BibitemOpen
  \bibfield  {author} {\bibinfo {author} {\bibfnamefont {K.}~\bibnamefont
  {Dholakia}}, \bibinfo {author} {\bibfnamefont {B.~W.}\ \bibnamefont
  {Drinkwater}},\ and\ \bibinfo {author} {\bibfnamefont {M.}~\bibnamefont
  {Ritsch-Marte}},\ }\bibfield  {title} {\bibinfo {title} {{Comparing acoustic
  and optical forces for biomedical research}},\ }\href
  {https://doi.org/10.1038/s42254-020-0215-3} {\bibfield  {journal} {\bibinfo
  {journal} {Nat. Rev. Phys.}\ }\textbf {\bibinfo {volume} {2}},\ \bibinfo
  {pages} {480} (\bibinfo {year} {2020})}\BibitemShut {NoStop}%
\bibitem [{\citenamefont {Fazal}\ and\ \citenamefont
  {Block}(2011)}]{Fazal2011Jun}%
  \BibitemOpen
  \bibfield  {author} {\bibinfo {author} {\bibfnamefont {F.~M.}\ \bibnamefont
  {Fazal}}\ and\ \bibinfo {author} {\bibfnamefont {S.~M.}\ \bibnamefont
  {Block}},\ }\bibfield  {title} {\bibinfo {title} {{Optical tweezers study
  life under tension}},\ }\href {https://doi.org/10.1038/nphoton.2011.100}
  {\bibfield  {journal} {\bibinfo  {journal} {Nat. Photonics}\ }\textbf
  {\bibinfo {volume} {5}},\ \bibinfo {pages} {318} (\bibinfo {year}
  {2011})}\BibitemShut {NoStop}%
\bibitem [{\citenamefont {Kaufman}\ and\ \citenamefont
  {Ni}(2021)}]{Kaufman2021Dec}%
  \BibitemOpen
  \bibfield  {author} {\bibinfo {author} {\bibfnamefont {A.~M.}\ \bibnamefont
  {Kaufman}}\ and\ \bibinfo {author} {\bibfnamefont {K.-K.}\ \bibnamefont
  {Ni}},\ }\bibfield  {title} {\bibinfo {title} {{Quantum science with optical
  tweezer arrays of ultracold atoms and molecules}},\ }\href
  {https://doi.org/10.1038/s41567-021-01357-2} {\bibfield  {journal} {\bibinfo
  {journal} {Nat. Phys.}\ }\textbf {\bibinfo {volume} {17}},\ \bibinfo {pages}
  {1324} (\bibinfo {year} {2021})}\BibitemShut {NoStop}%
\bibitem [{\citenamefont {Kim}\ \emph {et~al.}(2019)\citenamefont {Kim},
  \citenamefont {Chang}, \citenamefont {Fields}, \citenamefont {Chen},\ and\
  \citenamefont {Hung}}]{Kim2019Apr}%
  \BibitemOpen
  \bibfield  {author} {\bibinfo {author} {\bibfnamefont {M.~E.}\ \bibnamefont
  {Kim}}, \bibinfo {author} {\bibfnamefont {T.-H.}\ \bibnamefont {Chang}},
  \bibinfo {author} {\bibfnamefont {B.~M.}\ \bibnamefont {Fields}}, \bibinfo
  {author} {\bibfnamefont {C.-A.}\ \bibnamefont {Chen}},\ and\ \bibinfo
  {author} {\bibfnamefont {C.-L.}\ \bibnamefont {Hung}},\ }\bibfield  {title}
  {\bibinfo {title} {{Trapping single atoms on a nanophotonic circuit with
  configurable tweezer lattices}},\ }\href
  {https://doi.org/10.1038/s41467-019-09635-7} {\bibfield  {journal} {\bibinfo
  {journal} {Nat. Commun.}\ }\textbf {\bibinfo {volume} {10}},\ \bibinfo
  {pages} {1} (\bibinfo {year} {2019})}\BibitemShut {NoStop}%
\bibitem [{\citenamefont {Marag{\ifmmode\grave{o}\else\`{o}\fi}}\ \emph
  {et~al.}(2013)\citenamefont {Marag{\ifmmode\grave{o}\else\`{o}\fi}},
  \citenamefont {Jones}, \citenamefont {Gucciardi}, \citenamefont {Volpe},\
  and\ \citenamefont {Ferrari}}]{Marago2013Nov}%
  \BibitemOpen
  \bibfield  {author} {\bibinfo {author} {\bibfnamefont {O.~M.}\ \bibnamefont
  {Marag{\ifmmode\grave{o}\else\`{o}\fi}}}, \bibinfo {author} {\bibfnamefont
  {P.~H.}\ \bibnamefont {Jones}}, \bibinfo {author} {\bibfnamefont {P.~G.}\
  \bibnamefont {Gucciardi}}, \bibinfo {author} {\bibfnamefont {G.}~\bibnamefont
  {Volpe}},\ and\ \bibinfo {author} {\bibfnamefont {A.~C.}\ \bibnamefont
  {Ferrari}},\ }\bibfield  {title} {\bibinfo {title} {{Optical trapping and
  manipulation of nanostructures}},\ }\href
  {https://doi.org/10.1038/nnano.2013.208} {\bibfield  {journal} {\bibinfo
  {journal} {Nat. Nanotechnol.}\ }\textbf {\bibinfo {volume} {8}},\ \bibinfo
  {pages} {807} (\bibinfo {year} {2013})}\BibitemShut {NoStop}%
\bibitem [{\citenamefont {Shi}\ \emph {et~al.}(2022)\citenamefont {Shi},
  \citenamefont {Song}, \citenamefont {Toftul}, \citenamefont {Zhu},
  \citenamefont {Yu}, \citenamefont {Zhu}, \citenamefont {Tsai}, \citenamefont
  {Kivshar},\ and\ \citenamefont {Liu}}]{Shi2022Sep}%
  \BibitemOpen
  \bibfield  {author} {\bibinfo {author} {\bibfnamefont {Y.}~\bibnamefont
  {Shi}}, \bibinfo {author} {\bibfnamefont {Q.}~\bibnamefont {Song}}, \bibinfo
  {author} {\bibfnamefont {I.}~\bibnamefont {Toftul}}, \bibinfo {author}
  {\bibfnamefont {T.}~\bibnamefont {Zhu}}, \bibinfo {author} {\bibfnamefont
  {Y.}~\bibnamefont {Yu}}, \bibinfo {author} {\bibfnamefont {W.}~\bibnamefont
  {Zhu}}, \bibinfo {author} {\bibfnamefont {D.~P.}\ \bibnamefont {Tsai}},
  \bibinfo {author} {\bibfnamefont {Y.}~\bibnamefont {Kivshar}},\ and\ \bibinfo
  {author} {\bibfnamefont {A.~Q.}\ \bibnamefont {Liu}},\ }\bibfield  {title}
  {\bibinfo {title} {{Optical manipulation with metamaterial structures}},\
  }\href {https://doi.org/10.1063/5.0091280} {\bibfield  {journal} {\bibinfo
  {journal} {Appl. Phys. Rev.}\ }\textbf {\bibinfo {volume} {9}},\ \bibinfo
  {pages} {031303} (\bibinfo {year} {2022})}\BibitemShut {NoStop}%
\bibitem [{\citenamefont {Kostina}\ \emph {et~al.}(2019)\citenamefont
  {Kostina}, \citenamefont {Petrov}, \citenamefont {Ivinskaya}, \citenamefont
  {Sukhov}, \citenamefont {Bogdanov}, \citenamefont {Toftul}, \citenamefont
  {Nieto-Vesperinas}, \citenamefont {Ginzburg},\ and\ \citenamefont
  {Shalin}}]{Kostina2019Mar}%
  \BibitemOpen
  \bibfield  {author} {\bibinfo {author} {\bibfnamefont {N.}~\bibnamefont
  {Kostina}}, \bibinfo {author} {\bibfnamefont {M.}~\bibnamefont {Petrov}},
  \bibinfo {author} {\bibfnamefont {A.}~\bibnamefont {Ivinskaya}}, \bibinfo
  {author} {\bibfnamefont {S.}~\bibnamefont {Sukhov}}, \bibinfo {author}
  {\bibfnamefont {A.}~\bibnamefont {Bogdanov}}, \bibinfo {author}
  {\bibfnamefont {I.}~\bibnamefont {Toftul}}, \bibinfo {author} {\bibfnamefont
  {M.}~\bibnamefont {Nieto-Vesperinas}}, \bibinfo {author} {\bibfnamefont
  {P.}~\bibnamefont {Ginzburg}},\ and\ \bibinfo {author} {\bibfnamefont
  {A.}~\bibnamefont {Shalin}},\ }\bibfield  {title} {\bibinfo {title} {{Optical
  binding via surface plasmon polariton interference}},\ }\href
  {https://doi.org/10.1103/PhysRevB.99.125416} {\bibfield  {journal} {\bibinfo
  {journal} {Phys. Rev. B}\ }\textbf {\bibinfo {volume} {99}},\ \bibinfo
  {pages} {125416} (\bibinfo {year} {2019})}\BibitemShut {NoStop}%
\bibitem [{\citenamefont {Tkachenko}\ \emph {et~al.}(2020)\citenamefont
  {Tkachenko}, \citenamefont {Tkachenko}, \citenamefont {Toftul}, \citenamefont
  {Esporlas}, \citenamefont {Maimaiti}, \citenamefont {Maimaiti}, \citenamefont
  {Maimaiti}, \citenamefont {Le~Kien}, \citenamefont {Truong}, \citenamefont
  {Truong}, \citenamefont {Chormaic}, \citenamefont {Chormaic},\ and\
  \citenamefont {Chormaic}}]{Tkachenko2020Jan}%
  \BibitemOpen
  \bibfield  {author} {\bibinfo {author} {\bibfnamefont {G.}~\bibnamefont
  {Tkachenko}}, \bibinfo {author} {\bibfnamefont {G.}~\bibnamefont
  {Tkachenko}}, \bibinfo {author} {\bibfnamefont {I.}~\bibnamefont {Toftul}},
  \bibinfo {author} {\bibfnamefont {C.}~\bibnamefont {Esporlas}}, \bibinfo
  {author} {\bibfnamefont {A.}~\bibnamefont {Maimaiti}}, \bibinfo {author}
  {\bibfnamefont {A.}~\bibnamefont {Maimaiti}}, \bibinfo {author}
  {\bibfnamefont {A.}~\bibnamefont {Maimaiti}}, \bibinfo {author}
  {\bibfnamefont {F.}~\bibnamefont {Le~Kien}}, \bibinfo {author} {\bibfnamefont
  {V.~G.}\ \bibnamefont {Truong}}, \bibinfo {author} {\bibfnamefont {V.~G.}\
  \bibnamefont {Truong}}, \bibinfo {author} {\bibfnamefont {S.~N.}\
  \bibnamefont {Chormaic}}, \bibinfo {author} {\bibfnamefont {S.~N.}\
  \bibnamefont {Chormaic}},\ and\ \bibinfo {author} {\bibfnamefont {S.~N.}\
  \bibnamefont {Chormaic}},\ }\bibfield  {title} {\bibinfo {title}
  {{Light-induced rotation of dielectric microparticles around an optical
  nanofiber}},\ }\href {https://doi.org/10.1364/OPTICA.374441} {\bibfield
  {journal} {\bibinfo  {journal} {Optica}\ }\textbf {\bibinfo {volume} {7}},\
  \bibinfo {pages} {59} (\bibinfo {year} {2020})}\BibitemShut {NoStop}%
\bibitem [{\citenamefont {Toftul}\ \emph {et~al.}(2020)\citenamefont {Toftul},
  \citenamefont {Kornovan},\ and\ \citenamefont {Petrov}}]{Toftul2020Jan}%
  \BibitemOpen
  \bibfield  {author} {\bibinfo {author} {\bibfnamefont {I.~D.}\ \bibnamefont
  {Toftul}}, \bibinfo {author} {\bibfnamefont {D.~F.}\ \bibnamefont
  {Kornovan}},\ and\ \bibinfo {author} {\bibfnamefont {M.~I.}\ \bibnamefont
  {Petrov}},\ }\bibfield  {title} {\bibinfo {title} {{Self-Trapped Nanoparticle
  Binding via Waveguide Mode}},\ }\href
  {https://doi.org/10.1021/acsphotonics.9b01157} {\bibfield  {journal}
  {\bibinfo  {journal} {ACS Photonics}\ }\textbf {\bibinfo {volume} {7}},\
  \bibinfo {pages} {114} (\bibinfo {year} {2020})}\BibitemShut {NoStop}%
\bibitem [{\citenamefont
  {Jackson}(1998)}]{jackson1998ClassicalElectrodynamics}%
  \BibitemOpen
  \bibfield  {author} {\bibinfo {author} {\bibfnamefont {J.~D.}\ \bibnamefont
  {Jackson}},\ }\href@noop {} {\emph {\bibinfo {title} {Classical
  {{Electrodynamics}}}}},\ Vol.~\bibinfo {volume} {1}\ (\bibinfo {year}
  {1998})\BibitemShut {NoStop}%
\bibitem [{\citenamefont {Novotny}\ and\ \citenamefont
  {Hetch}(2010)}]{novotny2010PrinciplesNanoOptics}%
  \BibitemOpen
  \bibfield  {author} {\bibinfo {author} {\bibfnamefont {L.}~\bibnamefont
  {Novotny}}\ and\ \bibinfo {author} {\bibfnamefont {B.}~\bibnamefont
  {Hetch}},\ }\href@noop {} {\emph {\bibinfo {title} {Principles of
  {{Nano}}-{{Optics}}}}},\ Vol.~\bibinfo {volume} {1}\ (\bibinfo {year}
  {2010})\BibitemShut {NoStop}%
\bibitem [{\citenamefont {L.~Allen}(2014)}]{L.Allen2014Apr}%
  \BibitemOpen
  \bibfield  {author} {\bibinfo {author} {\bibfnamefont {S.~M.~B.}\
  \bibnamefont {L.~Allen}},\ }\href {https://doi.org/10.1201/9781482269017}
  {\emph {\bibinfo {title} {{Optical Angular Momentum}}}}\ (\bibinfo
  {publisher} {Taylor {\&} Francis},\ \bibinfo {address} {Andover, England,
  UK},\ \bibinfo {year} {2014})\BibitemShut {NoStop}%
\bibitem [{\citenamefont {Ye}\ and\ \citenamefont {Lin}(2017)}]{Ye2017May}%
  \BibitemOpen
  \bibfield  {author} {\bibinfo {author} {\bibfnamefont {Q.}~\bibnamefont
  {Ye}}\ and\ \bibinfo {author} {\bibfnamefont {H.}~\bibnamefont {Lin}},\
  }\bibfield  {title} {\bibinfo {title} {{On deriving the Maxwell stress tensor
  method for calculating the optical force and torque on an object in harmonic
  electromagnetic fields}},\ }\href {https://doi.org/10.1088/1361-6404/aa6e1d}
  {\bibfield  {journal} {\bibinfo  {journal} {Eur. J. Phys.}\ }\textbf
  {\bibinfo {volume} {38}},\ \bibinfo {pages} {045202} (\bibinfo {year}
  {2017})}\BibitemShut {NoStop}%
\bibitem [{\citenamefont {Brasselet}\ and\ \citenamefont
  {Juodkazis}(2009)}]{Brasselet2009Jun}%
  \BibitemOpen
  \bibfield  {author} {\bibinfo {author} {\bibfnamefont {E.}~\bibnamefont
  {Brasselet}}\ and\ \bibinfo {author} {\bibfnamefont {S.}~\bibnamefont
  {Juodkazis}},\ }\bibfield  {title} {\bibinfo {title} {{Optical angular
  manipulation of liquid crystal droplets in laser tweezers}},\ }\href
  {https://doi.org/10.1142/S0218863509004580} {\bibfield  {journal} {\bibinfo
  {journal} {J. Nonlinear Opt. Phys. Mater.}\ }\textbf {\bibinfo {volume}
  {18}},\ \bibinfo {pages} {167} (\bibinfo {year} {2009})}\BibitemShut
  {NoStop}%
\bibitem [{\citenamefont {Friese}\ \emph {et~al.}(1998)\citenamefont {Friese},
  \citenamefont {Nieminen}, \citenamefont {Heckenberg},\ and\ \citenamefont
  {Rubinsztein-Dunlop}}]{Friese1998Jul}%
  \BibitemOpen
  \bibfield  {author} {\bibinfo {author} {\bibfnamefont {M.~E.~J.}\
  \bibnamefont {Friese}}, \bibinfo {author} {\bibfnamefont {T.~A.}\
  \bibnamefont {Nieminen}}, \bibinfo {author} {\bibfnamefont {N.~R.}\
  \bibnamefont {Heckenberg}},\ and\ \bibinfo {author} {\bibfnamefont
  {H.}~\bibnamefont {Rubinsztein-Dunlop}},\ }\bibfield  {title} {\bibinfo
  {title} {{Optical alignment and spinning of laser-trapped microscopic
  particles}},\ }\href {https://doi.org/10.1038/28566} {\bibfield  {journal}
  {\bibinfo  {journal} {Nature}\ }\textbf {\bibinfo {volume} {394}},\ \bibinfo
  {pages} {348} (\bibinfo {year} {1998})}\BibitemShut {NoStop}%
\bibitem [{\citenamefont {Simpson}\ \emph {et~al.}(2007)\citenamefont
  {Simpson}, \citenamefont {Benito},\ and\ \citenamefont
  {Hanna}}]{Simpson2007Oct}%
  \BibitemOpen
  \bibfield  {author} {\bibinfo {author} {\bibfnamefont {S.~H.}\ \bibnamefont
  {Simpson}}, \bibinfo {author} {\bibfnamefont {D.~C.}\ \bibnamefont
  {Benito}},\ and\ \bibinfo {author} {\bibfnamefont {S.}~\bibnamefont
  {Hanna}},\ }\bibfield  {title} {\bibinfo {title} {{Polarization-induced
  torque in optical traps}},\ }\href
  {https://doi.org/10.1103/PhysRevA.76.043408} {\bibfield  {journal} {\bibinfo
  {journal} {Phys. Rev. A}\ }\textbf {\bibinfo {volume} {76}},\ \bibinfo
  {pages} {043408} (\bibinfo {year} {2007})}\BibitemShut {NoStop}%
\bibitem [{\citenamefont {Trojek}\ \emph {et~al.}(2012)\citenamefont {Trojek},
  \citenamefont {Chv{\ifmmode\acute{a}\else\'{a}\fi}tal},\ and\ \citenamefont
  {Zem{\ifmmode\acute{a}\else\'{a}\fi}nek}}]{Trojek2012Jul}%
  \BibitemOpen
  \bibfield  {author} {\bibinfo {author} {\bibfnamefont {J.}~\bibnamefont
  {Trojek}}, \bibinfo {author} {\bibfnamefont {L.}~\bibnamefont
  {Chv{\ifmmode\acute{a}\else\'{a}\fi}tal}},\ and\ \bibinfo {author}
  {\bibfnamefont {P.}~\bibnamefont {Zem{\ifmmode\acute{a}\else\'{a}\fi}nek}},\
  }\bibfield  {title} {\bibinfo {title} {{Optical alignment and confinement of
  an ellipsoidal nanorod in optical tweezers: a theoretical study}},\ }\href
  {https://doi.org/10.1364/JOSAA.29.001224} {\bibfield  {journal} {\bibinfo
  {journal} {J. Opt. Soc. Am. A, JOSAA}\ }\textbf {\bibinfo {volume} {29}},\
  \bibinfo {pages} {1224} (\bibinfo {year} {2012})}\BibitemShut {NoStop}%
\bibitem [{\citenamefont {Marston}\ and\ \citenamefont
  {Crichton}(1984)}]{Marston1984Nov}%
  \BibitemOpen
  \bibfield  {author} {\bibinfo {author} {\bibfnamefont {P.~L.}\ \bibnamefont
  {Marston}}\ and\ \bibinfo {author} {\bibfnamefont {J.~H.}\ \bibnamefont
  {Crichton}},\ }\bibfield  {title} {\bibinfo {title} {{Radiation torque on a
  sphere caused by a circularly-polarized electromagnetic wave}},\ }\href
  {https://doi.org/10.1103/PhysRevA.30.2508} {\bibfield  {journal} {\bibinfo
  {journal} {Phys. Rev. A}\ }\textbf {\bibinfo {volume} {30}},\ \bibinfo
  {pages} {2508} (\bibinfo {year} {1984})}\BibitemShut {NoStop}%
\bibitem [{\citenamefont {Canaguier-Durand}\ \emph {et~al.}(2013)\citenamefont
  {Canaguier-Durand}, \citenamefont {Cuche}, \citenamefont {Genet},\ and\
  \citenamefont {Ebbesen}}]{Canaguier-Durand2013Sep}%
  \BibitemOpen
  \bibfield  {author} {\bibinfo {author} {\bibfnamefont {A.}~\bibnamefont
  {Canaguier-Durand}}, \bibinfo {author} {\bibfnamefont {A.}~\bibnamefont
  {Cuche}}, \bibinfo {author} {\bibfnamefont {C.}~\bibnamefont {Genet}},\ and\
  \bibinfo {author} {\bibfnamefont {T.~W.}\ \bibnamefont {Ebbesen}},\
  }\bibfield  {title} {\bibinfo {title} {{Force and torque on an electric
  dipole by spinning light fields}},\ }\href
  {https://doi.org/10.1103/PhysRevA.88.033831} {\bibfield  {journal} {\bibinfo
  {journal} {Phys. Rev. A}\ }\textbf {\bibinfo {volume} {88}},\ \bibinfo
  {pages} {033831} (\bibinfo {year} {2013})}\BibitemShut {NoStop}%
\bibitem [{\citenamefont {Chen}\ \emph {et~al.}(2014)\citenamefont {Chen},
  \citenamefont {Ng}, \citenamefont {Ding}, \citenamefont {Fung}, \citenamefont
  {Lin},\ and\ \citenamefont {Chan}}]{Chen2014Sep}%
  \BibitemOpen
  \bibfield  {author} {\bibinfo {author} {\bibfnamefont {J.}~\bibnamefont
  {Chen}}, \bibinfo {author} {\bibfnamefont {J.}~\bibnamefont {Ng}}, \bibinfo
  {author} {\bibfnamefont {K.}~\bibnamefont {Ding}}, \bibinfo {author}
  {\bibfnamefont {K.~H.}\ \bibnamefont {Fung}}, \bibinfo {author}
  {\bibfnamefont {Z.}~\bibnamefont {Lin}},\ and\ \bibinfo {author}
  {\bibfnamefont {C.~T.}\ \bibnamefont {Chan}},\ }\bibfield  {title} {\bibinfo
  {title} {{Negative Optical Torque - Scientific Reports}},\ }\href
  {https://doi.org/10.1038/srep06386} {\bibfield  {journal} {\bibinfo
  {journal} {Sci. Rep.}\ }\textbf {\bibinfo {volume} {4}},\ \bibinfo {pages}
  {1} (\bibinfo {year} {2014})}\BibitemShut {NoStop}%
\bibitem [{\citenamefont {Mitri}(2016)}]{Mitri2016Oct}%
  \BibitemOpen
  \bibfield  {author} {\bibinfo {author} {\bibfnamefont {F.~G.}\ \bibnamefont
  {Mitri}},\ }\bibfield  {title} {\bibinfo {title} {{Negative optical spin
  torque wrench of a non-diffracting non-paraxial fractional Bessel vortex
  beam}},\ }\href {https://doi.org/10.1016/j.jqsrt.2016.05.033} {\bibfield
  {journal} {\bibinfo  {journal} {J. Quant. Spectrosc. Radiat. Transfer}\
  }\textbf {\bibinfo {volume} {182}},\ \bibinfo {pages} {172} (\bibinfo {year}
  {2016})}\BibitemShut {NoStop}%
\bibitem [{\citenamefont {Nieto-Vesperinas}(2015)}]{Nieto-Vesperinas2015Jul}%
  \BibitemOpen
  \bibfield  {author} {\bibinfo {author} {\bibfnamefont {M.}~\bibnamefont
  {Nieto-Vesperinas}},\ }\bibfield  {title} {\bibinfo {title} {{Optical torque
  on small bi-isotropic particles}},\ }\href
  {https://doi.org/10.1364/OL.40.003021} {\bibfield  {journal} {\bibinfo
  {journal} {Opt. Lett.}\ }\textbf {\bibinfo {volume} {40}},\ \bibinfo {pages}
  {3021} (\bibinfo {year} {2015})}\BibitemShut {NoStop}%
\bibitem [{\citenamefont {Han}\ \emph {et~al.}(2018)\citenamefont {Han},
  \citenamefont {Parker}, \citenamefont {Yifat}, \citenamefont {Peterson},
  \citenamefont {Gray}, \citenamefont {Scherer},\ and\ \citenamefont
  {Yan}}]{Han2018Nov}%
  \BibitemOpen
  \bibfield  {author} {\bibinfo {author} {\bibfnamefont {F.}~\bibnamefont
  {Han}}, \bibinfo {author} {\bibfnamefont {J.~A.}\ \bibnamefont {Parker}},
  \bibinfo {author} {\bibfnamefont {Y.}~\bibnamefont {Yifat}}, \bibinfo
  {author} {\bibfnamefont {C.}~\bibnamefont {Peterson}}, \bibinfo {author}
  {\bibfnamefont {S.~K.}\ \bibnamefont {Gray}}, \bibinfo {author}
  {\bibfnamefont {N.~F.}\ \bibnamefont {Scherer}},\ and\ \bibinfo {author}
  {\bibfnamefont {Z.}~\bibnamefont {Yan}},\ }\bibfield  {title} {\bibinfo
  {title} {{Crossover from positive to negative optical torque in mesoscale
  optical matter}},\ }\href {https://doi.org/10.1038/s41467-018-07376-7}
  {\bibfield  {journal} {\bibinfo  {journal} {Nat. Commun.}\ }\textbf {\bibinfo
  {volume} {9}},\ \bibinfo {pages} {1} (\bibinfo {year} {2018})}\BibitemShut
  {NoStop}%
\bibitem [{\citenamefont {Sule}\ \emph {et~al.}(2017)\citenamefont {Sule},
  \citenamefont {Yifat}, \citenamefont {Gray},\ and\ \citenamefont
  {Scherer}}]{Sule2017Nov}%
  \BibitemOpen
  \bibfield  {author} {\bibinfo {author} {\bibfnamefont {N.}~\bibnamefont
  {Sule}}, \bibinfo {author} {\bibfnamefont {Y.}~\bibnamefont {Yifat}},
  \bibinfo {author} {\bibfnamefont {S.~K.}\ \bibnamefont {Gray}},\ and\
  \bibinfo {author} {\bibfnamefont {N.~F.}\ \bibnamefont {Scherer}},\
  }\bibfield  {title} {\bibinfo {title} {{Rotation and Negative Torque in
  Electrodynamically Bound Nanoparticle Dimers}},\ }\href
  {https://doi.org/10.1021/acs.nanolett.7b02196} {\bibfield  {journal}
  {\bibinfo  {journal} {Nano Lett.}\ }\textbf {\bibinfo {volume} {17}},\
  \bibinfo {pages} {6548} (\bibinfo {year} {2017})}\BibitemShut {NoStop}%
\bibitem [{\citenamefont {Diniz}\ \emph {et~al.}(2019)\citenamefont {Diniz},
  \citenamefont {Dutra}, \citenamefont {Pires}, \citenamefont {Viana},
  \citenamefont {Nussenzveig},\ and\ \citenamefont {Neto}}]{Diniz2019Mar}%
  \BibitemOpen
  \bibfield  {author} {\bibinfo {author} {\bibfnamefont {K.}~\bibnamefont
  {Diniz}}, \bibinfo {author} {\bibfnamefont {R.~S.}\ \bibnamefont {Dutra}},
  \bibinfo {author} {\bibfnamefont {L.~B.}\ \bibnamefont {Pires}}, \bibinfo
  {author} {\bibfnamefont {N.~B.}\ \bibnamefont {Viana}}, \bibinfo {author}
  {\bibfnamefont {H.~M.}\ \bibnamefont {Nussenzveig}},\ and\ \bibinfo {author}
  {\bibfnamefont {P.~A.~M.}\ \bibnamefont {Neto}},\ }\bibfield  {title}
  {\bibinfo {title} {{Negative optical torque on a microsphere in optical
  tweezers}},\ }\href {https://doi.org/10.1364/OE.27.005905} {\bibfield
  {journal} {\bibinfo  {journal} {Opt. Express}\ }\textbf {\bibinfo {volume}
  {27}},\ \bibinfo {pages} {5905} (\bibinfo {year} {2019})}\BibitemShut
  {NoStop}%
\bibitem [{\citenamefont {Hakobyan}\ and\ \citenamefont
  {Brasselet}(2014)}]{Hakobyan2014Aug}%
  \BibitemOpen
  \bibfield  {author} {\bibinfo {author} {\bibfnamefont {D.}~\bibnamefont
  {Hakobyan}}\ and\ \bibinfo {author} {\bibfnamefont {E.}~\bibnamefont
  {Brasselet}},\ }\bibfield  {title} {\bibinfo {title} {{Left-handed optical
  radiation torque}},\ }\href {https://doi.org/10.1038/nphoton.2014.142}
  {\bibfield  {journal} {\bibinfo  {journal} {Nat. Photonics}\ }\textbf
  {\bibinfo {volume} {8}},\ \bibinfo {pages} {610} (\bibinfo {year}
  {2014})}\BibitemShut {NoStop}%
\bibitem [{\citenamefont {Zograf}\ \emph {et~al.}(2021)\citenamefont {Zograf},
  \citenamefont {Zograf}, \citenamefont {Petrov}, \citenamefont {Makarov},
  \citenamefont {Kivshar},\ and\ \citenamefont {Kivshar}}]{Zograf2021Sep}%
  \BibitemOpen
  \bibfield  {author} {\bibinfo {author} {\bibfnamefont {G.~P.}\ \bibnamefont
  {Zograf}}, \bibinfo {author} {\bibfnamefont {G.~P.}\ \bibnamefont {Zograf}},
  \bibinfo {author} {\bibfnamefont {M.~I.}\ \bibnamefont {Petrov}}, \bibinfo
  {author} {\bibfnamefont {S.~V.}\ \bibnamefont {Makarov}}, \bibinfo {author}
  {\bibfnamefont {Y.~S.}\ \bibnamefont {Kivshar}},\ and\ \bibinfo {author}
  {\bibfnamefont {Y.~S.}\ \bibnamefont {Kivshar}},\ }\bibfield  {title}
  {\bibinfo {title} {{All-dielectric thermonanophotonics}},\ }\href
  {https://doi.org/10.1364/AOP.426047} {\bibfield  {journal} {\bibinfo
  {journal} {Adv. Opt. Photonics}\ }\textbf {\bibinfo {volume} {13}},\ \bibinfo
  {pages} {643} (\bibinfo {year} {2021})}\BibitemShut {NoStop}%
\bibitem [{\citenamefont {Baranov}\ \emph {et~al.}(2017)\citenamefont
  {Baranov}, \citenamefont {Zuev}, \citenamefont {Lepeshov}, \citenamefont
  {Kotov}, \citenamefont {Krasnok}, \citenamefont {Evlyukhin},\ and\
  \citenamefont {Chichkov}}]{Baranov2017Jul}%
  \BibitemOpen
  \bibfield  {author} {\bibinfo {author} {\bibfnamefont {D.~G.}\ \bibnamefont
  {Baranov}}, \bibinfo {author} {\bibfnamefont {D.~A.}\ \bibnamefont {Zuev}},
  \bibinfo {author} {\bibfnamefont {S.~I.}\ \bibnamefont {Lepeshov}}, \bibinfo
  {author} {\bibfnamefont {O.~V.}\ \bibnamefont {Kotov}}, \bibinfo {author}
  {\bibfnamefont {A.~E.}\ \bibnamefont {Krasnok}}, \bibinfo {author}
  {\bibfnamefont {A.~B.}\ \bibnamefont {Evlyukhin}},\ and\ \bibinfo {author}
  {\bibfnamefont {B.~N.}\ \bibnamefont {Chichkov}},\ }\bibfield  {title}
  {\bibinfo {title} {{All-dielectric nanophotonics: the quest for better
  materials and fabrication techniques}},\ }\href
  {https://doi.org/10.1364/OPTICA.4.000814} {\bibfield  {journal} {\bibinfo
  {journal} {Optica}\ }\textbf {\bibinfo {volume} {4}},\ \bibinfo {pages} {814}
  (\bibinfo {year} {2017})}\BibitemShut {NoStop}%
\bibitem [{\citenamefont {Kivshar}(2018)}]{Kivshar2018Mar}%
  \BibitemOpen
  \bibfield  {author} {\bibinfo {author} {\bibfnamefont {Y.}~\bibnamefont
  {Kivshar}},\ }\bibfield  {title} {\bibinfo {title} {{All-dielectric
  meta-optics and non-linear nanophotonics}},\ }\href
  {https://doi.org/10.1093/nsr/nwy017} {\bibfield  {journal} {\bibinfo
  {journal} {Natl. Sci. Rev.}\ }\textbf {\bibinfo {volume} {5}},\ \bibinfo
  {pages} {144} (\bibinfo {year} {2018})}\BibitemShut {NoStop}%
\bibitem [{\citenamefont {Krasnok}\ \emph {et~al.}(2015)\citenamefont
  {Krasnok}, \citenamefont {Makarov}, \citenamefont {Petrov}, \citenamefont
  {Savelev}, \citenamefont {Belov},\ and\ \citenamefont
  {Kivshar}}]{Krasnok2015May}%
  \BibitemOpen
  \bibfield  {author} {\bibinfo {author} {\bibfnamefont {A.}~\bibnamefont
  {Krasnok}}, \bibinfo {author} {\bibfnamefont {S.}~\bibnamefont {Makarov}},
  \bibinfo {author} {\bibfnamefont {M.}~\bibnamefont {Petrov}}, \bibinfo
  {author} {\bibfnamefont {R.}~\bibnamefont {Savelev}}, \bibinfo {author}
  {\bibfnamefont {P.}~\bibnamefont {Belov}},\ and\ \bibinfo {author}
  {\bibfnamefont {Y.}~\bibnamefont {Kivshar}},\ }\bibfield  {title} {\bibinfo
  {title} {{Towards all-dielectric metamaterials and nanophotonics}},\ }in\
  \href {https://doi.org/10.1117/12.2176880} {\emph {\bibinfo {booktitle}
  {{Proceedings Volume 9502, Metamaterials X}}}},\ Vol.\ \bibinfo {volume}
  {9502}\ (\bibinfo  {publisher} {SPIE},\ \bibinfo {year} {2015})\ p.\ \bibinfo
  {pages} {950203}\BibitemShut {NoStop}%
\bibitem [{\citenamefont {Decker}\ and\ \citenamefont
  {Staude}(2016)}]{Decker2016Sep}%
  \BibitemOpen
  \bibfield  {author} {\bibinfo {author} {\bibfnamefont {M.}~\bibnamefont
  {Decker}}\ and\ \bibinfo {author} {\bibfnamefont {I.}~\bibnamefont
  {Staude}},\ }\bibfield  {title} {\bibinfo {title} {{Resonant dielectric
  nanostructures: a low-loss platform for functional nanophotonics}},\ }\href
  {https://doi.org/10.1088/2040-8978/18/10/103001} {\bibfield  {journal}
  {\bibinfo  {journal} {J. Opt.}\ }\textbf {\bibinfo {volume} {18}},\ \bibinfo
  {pages} {103001} (\bibinfo {year} {2016})}\BibitemShut {NoStop}%
\bibitem [{\citenamefont {Malmqvist}\ and\ \citenamefont
  {Hertz}(1995)}]{Malmqvist1995Jun}%
  \BibitemOpen
  \bibfield  {author} {\bibinfo {author} {\bibfnamefont {L.}~\bibnamefont
  {Malmqvist}}\ and\ \bibinfo {author} {\bibfnamefont {H.~M.}\ \bibnamefont
  {Hertz}},\ }\bibfield  {title} {\bibinfo {title} {{Second-harmonic generation
  in optically trapped nonlinear particles with pulsed lasers}},\ }\href
  {https://doi.org/10.1364/AO.34.003392} {\bibfield  {journal} {\bibinfo
  {journal} {Appl. Opt.}\ }\textbf {\bibinfo {volume} {34}},\ \bibinfo {pages}
  {3392} (\bibinfo {year} {1995})}\BibitemShut {NoStop}%
\bibitem [{\citenamefont {Sato}\ and\ \citenamefont
  {Inaba}(1994)}]{Sato1994Jul}%
  \BibitemOpen
  \bibfield  {author} {\bibinfo {author} {\bibfnamefont {S.}~\bibnamefont
  {Sato}}\ and\ \bibinfo {author} {\bibfnamefont {H.}~\bibnamefont {Inaba}},\
  }\bibfield  {title} {\bibinfo {title} {{Second-harmonic and sum-frequency
  generation from optically trapped KTiOPO4 microscopic particles by use of
  Nd:YAG and Ti:Al2O3 lasers}},\ }\href {https://doi.org/10.1364/OL.19.000927}
  {\bibfield  {journal} {\bibinfo  {journal} {Opt. Lett.}\ }\textbf {\bibinfo
  {volume} {19}},\ \bibinfo {pages} {927} (\bibinfo {year} {1994})}\BibitemShut
  {NoStop}%
\bibitem [{\citenamefont {Frizyuk}\ \emph {et~al.}(2019)\citenamefont
  {Frizyuk}, \citenamefont {Volkovskaya}, \citenamefont {Smirnova},
  \citenamefont {Poddubny},\ and\ \citenamefont {Petrov}}]{Frizyuk2019Feb}%
  \BibitemOpen
  \bibfield  {author} {\bibinfo {author} {\bibfnamefont {K.}~\bibnamefont
  {Frizyuk}}, \bibinfo {author} {\bibfnamefont {I.}~\bibnamefont
  {Volkovskaya}}, \bibinfo {author} {\bibfnamefont {D.}~\bibnamefont
  {Smirnova}}, \bibinfo {author} {\bibfnamefont {A.}~\bibnamefont {Poddubny}},\
  and\ \bibinfo {author} {\bibfnamefont {M.}~\bibnamefont {Petrov}},\
  }\bibfield  {title} {\bibinfo {title} {{Second-harmonic generation in
  Mie-resonant dielectric nanoparticles made of noncentrosymmetric
  materials}},\ }\href {https://doi.org/10.1103/PhysRevB.99.075425} {\bibfield
  {journal} {\bibinfo  {journal} {Phys. Rev. B}\ }\textbf {\bibinfo {volume}
  {99}},\ \bibinfo {pages} {075425} (\bibinfo {year} {2019})}\BibitemShut
  {NoStop}%
\bibitem [{\citenamefont {Koshelev}\ and\ \citenamefont
  {Kivshar}(2021)}]{Koshelev2021Jan}%
  \BibitemOpen
  \bibfield  {author} {\bibinfo {author} {\bibfnamefont {K.}~\bibnamefont
  {Koshelev}}\ and\ \bibinfo {author} {\bibfnamefont {Y.}~\bibnamefont
  {Kivshar}},\ }\bibfield  {title} {\bibinfo {title} {{Dielectric Resonant
  Metaphotonics}},\ }\href {https://doi.org/10.1021/acsphotonics.0c01315}
  {\bibfield  {journal} {\bibinfo  {journal} {ACS Photonics}\ }\textbf
  {\bibinfo {volume} {8}},\ \bibinfo {pages} {102} (\bibinfo {year}
  {2021})}\BibitemShut {NoStop}%
\bibitem [{\citenamefont {Carletti}\ \emph {et~al.}(2019)\citenamefont
  {Carletti}, \citenamefont {Kruk}, \citenamefont {Bogdanov}, \citenamefont
  {De~Angelis},\ and\ \citenamefont {Kivshar}}]{Carletti2019Sep}%
  \BibitemOpen
  \bibfield  {author} {\bibinfo {author} {\bibfnamefont {L.}~\bibnamefont
  {Carletti}}, \bibinfo {author} {\bibfnamefont {S.~S.}\ \bibnamefont {Kruk}},
  \bibinfo {author} {\bibfnamefont {A.~A.}\ \bibnamefont {Bogdanov}}, \bibinfo
  {author} {\bibfnamefont {C.}~\bibnamefont {De~Angelis}},\ and\ \bibinfo
  {author} {\bibfnamefont {Y.}~\bibnamefont {Kivshar}},\ }\bibfield  {title}
  {\bibinfo {title} {{High-harmonic generation at the nanoscale boosted by
  bound states in the continuum}},\ }\href
  {https://doi.org/10.1103/PhysRevResearch.1.023016} {\bibfield  {journal}
  {\bibinfo  {journal} {Phys. Rev. Res.}\ }\textbf {\bibinfo {volume} {1}},\
  \bibinfo {pages} {023016} (\bibinfo {year} {2019})}\BibitemShut {NoStop}%
\bibitem [{\citenamefont {Carletti}\ \emph {et~al.}(2018)\citenamefont
  {Carletti}, \citenamefont {Koshelev}, \citenamefont {De~Angelis},\ and\
  \citenamefont {Kivshar}}]{Carletti2018Jul}%
  \BibitemOpen
  \bibfield  {author} {\bibinfo {author} {\bibfnamefont {L.}~\bibnamefont
  {Carletti}}, \bibinfo {author} {\bibfnamefont {K.}~\bibnamefont {Koshelev}},
  \bibinfo {author} {\bibfnamefont {C.}~\bibnamefont {De~Angelis}},\ and\
  \bibinfo {author} {\bibfnamefont {Y.}~\bibnamefont {Kivshar}},\ }\bibfield
  {title} {\bibinfo {title} {{Giant Nonlinear Response at the Nanoscale Driven
  by Bound States in the Continuum}},\ }\href
  {https://doi.org/10.1103/PhysRevLett.121.033903} {\bibfield  {journal}
  {\bibinfo  {journal} {Phys. Rev. Lett.}\ }\textbf {\bibinfo {volume} {121}},\
  \bibinfo {pages} {033903} (\bibinfo {year} {2018})}\BibitemShut {NoStop}%
\bibitem [{\citenamefont {Bonacina}\ \emph {et~al.}(2020)\citenamefont
  {Bonacina}, \citenamefont {Brevet}, \citenamefont {Finazzi},\ and\
  \citenamefont {Celebrano}}]{Bonacina2020Jun}%
  \BibitemOpen
  \bibfield  {author} {\bibinfo {author} {\bibfnamefont {L.}~\bibnamefont
  {Bonacina}}, \bibinfo {author} {\bibfnamefont {P.-F.}\ \bibnamefont
  {Brevet}}, \bibinfo {author} {\bibfnamefont {M.}~\bibnamefont {Finazzi}},\
  and\ \bibinfo {author} {\bibfnamefont {M.}~\bibnamefont {Celebrano}},\
  }\bibfield  {title} {\bibinfo {title} {{Harmonic generation at the
  nanoscale}},\ }\href {https://doi.org/10.1063/5.0006093} {\bibfield
  {journal} {\bibinfo  {journal} {J. Appl. Phys.}\ }\textbf {\bibinfo {volume}
  {127}},\ \bibinfo {pages} {230901} (\bibinfo {year} {2020})}\BibitemShut
  {NoStop}%
\bibitem [{\citenamefont {Bliokh}\ \emph {et~al.}(2017)\citenamefont {Bliokh},
  \citenamefont {Bekshaev},\ and\ \citenamefont {Nori}}]{Bliokh2017Aug}%
  \BibitemOpen
  \bibfield  {author} {\bibinfo {author} {\bibfnamefont {K.~Y.}\ \bibnamefont
  {Bliokh}}, \bibinfo {author} {\bibfnamefont {A.~Y.}\ \bibnamefont
  {Bekshaev}},\ and\ \bibinfo {author} {\bibfnamefont {F.}~\bibnamefont
  {Nori}},\ }\bibfield  {title} {\bibinfo {title} {{Optical Momentum, Spin, and
  Angular Momentum in Dispersive Media}},\ }\href
  {https://doi.org/10.1103/PhysRevLett.119.073901} {\bibfield  {journal}
  {\bibinfo  {journal} {Phys. Rev. Lett.}\ }\textbf {\bibinfo {volume} {119}},\
  \bibinfo {pages} {073901} (\bibinfo {year} {2017})}\BibitemShut {NoStop}%
\bibitem [{\citenamefont {Ok}\ \emph {et~al.}(2006)\citenamefont {Ok},
  \citenamefont {Chi},\ and\ \citenamefont {Halasyamani}}]{Ok2006Jul}%
  \BibitemOpen
  \bibfield  {author} {\bibinfo {author} {\bibfnamefont {K.~M.}\ \bibnamefont
  {Ok}}, \bibinfo {author} {\bibfnamefont {E.~O.}\ \bibnamefont {Chi}},\ and\
  \bibinfo {author} {\bibfnamefont {P.~S.}\ \bibnamefont {Halasyamani}},\
  }\bibfield  {title} {\bibinfo {title} {{Bulk characterization methods for
  non-centrosymmetric materials: second-harmonic generation, piezoelectricity,
  pyroelectricity, and ferroelectricity}},\ }\href
  {https://doi.org/10.1039/B511119F} {\bibfield  {journal} {\bibinfo  {journal}
  {Chem. Soc. Rev.}\ }\textbf {\bibinfo {volume} {35}},\ \bibinfo {pages} {710}
  (\bibinfo {year} {2006})}\BibitemShut {NoStop}%
\bibitem [{SM()}]{SM}%
  \BibitemOpen
  \href@noop {} {\bibinfo {title} {{See Supplemental Material at \textit{[URL
  will be inserted by publisher]} for the optical torque numerical and
  analytical calculations details, explicit expressions for the coefficients
  $W^{\text E}_{mj}$ and $W^{\text H}_{mj}$, $\hat{\chi}^{(2)}$ tensor in
  cylindrical coordinates, estimations of the SHG efficiency and two- and
  three-photon absorption cross sections, rigid body rotational dynamics
  simulations, and detailed section on complex vector spherical harmonics,
  which includes
  Refs~\cite{bohren2008absorption,Doost2014Jul,Sehmi2020Jan,Lobanov2019Dec,Varshalovich1988Oct,Weisstein2003Nov,stratton2007ElectromagneticTheory,BibEntry2021Jun_comsol,jackson1998ClassicalElectrodynamics,dewitt1953UeberDrehimpulsMultipolstrahlung,novotny2010PrinciplesNanoOptics,bliokh2014ExtraordinaryMomentumSpin,li2014RadiationTorqueExerted,li2011CalculationRadiationForces,li2012CalculationRadiationForce,Barton1989,Barton1988,Li2014Dec,Li2012Jul,Pesce2020Dec,Borghese2006Oct,landau2013electrodynamics,Frizyuk2019Feb,toftul_github_rotation,Bergfeld2003Jan,Koshelev2020Jan,
  koshelev2022advanced,Doost2014Jul,kristensson2014spherical,Weiss2018Aug,Muljarov2018May,Perrin1934Oct,cantor1980biophysical,koenig1975brownian,landau2013fluid,Coutsias2004}
  }}}\BibitemShut {NoStop}%
\bibitem [{\citenamefont {Hurlbut}\ \emph {et~al.}(2007)\citenamefont
  {Hurlbut}, \citenamefont {Lee}, \citenamefont {Vodopyanov}, \citenamefont
  {Kuo},\ and\ \citenamefont {Fejer}}]{Hurlbut2007Mar}%
  \BibitemOpen
  \bibfield  {author} {\bibinfo {author} {\bibfnamefont {W.~C.}\ \bibnamefont
  {Hurlbut}}, \bibinfo {author} {\bibfnamefont {Y.-S.}\ \bibnamefont {Lee}},
  \bibinfo {author} {\bibfnamefont {K.~L.}\ \bibnamefont {Vodopyanov}},
  \bibinfo {author} {\bibfnamefont {P.~S.}\ \bibnamefont {Kuo}},\ and\ \bibinfo
  {author} {\bibfnamefont {M.~M.}\ \bibnamefont {Fejer}},\ }\bibfield  {title}
  {\bibinfo {title} {{Multiphoton absorption and nonlinear refraction of GaAs
  in the mid-infrared}},\ }\href {https://doi.org/10.1364/OL.32.000668}
  {\bibfield  {journal} {\bibinfo  {journal} {Opt. Lett.}\ }\textbf {\bibinfo
  {volume} {32}},\ \bibinfo {pages} {668} (\bibinfo {year} {2007})}\BibitemShut
  {NoStop}%
\bibitem [{\citenamefont {Griffiths}(2005)}]{griffiths2005introduction}%
  \BibitemOpen
  \bibfield  {author} {\bibinfo {author} {\bibfnamefont {D.~J.}\ \bibnamefont
  {Griffiths}},\ }\href@noop {} {\emph {\bibinfo {title} {Introduction to
  electrodynamics}}}\ (\bibinfo  {publisher} {American Association of Physics
  Teachers},\ \bibinfo {year} {2005})\BibitemShut {NoStop}%
\bibitem [{\citenamefont {Mun}\ \emph {et~al.}(2020)\citenamefont {Mun},
  \citenamefont {Kim}, \citenamefont {Yang}, \citenamefont {Badloe},
  \citenamefont {Ni}, \citenamefont {Chen}, \citenamefont {Qiu},\ and\
  \citenamefont {Rho}}]{Mun2020Sep}%
  \BibitemOpen
  \bibfield  {author} {\bibinfo {author} {\bibfnamefont {J.}~\bibnamefont
  {Mun}}, \bibinfo {author} {\bibfnamefont {M.}~\bibnamefont {Kim}}, \bibinfo
  {author} {\bibfnamefont {Y.}~\bibnamefont {Yang}}, \bibinfo {author}
  {\bibfnamefont {T.}~\bibnamefont {Badloe}}, \bibinfo {author} {\bibfnamefont
  {J.}~\bibnamefont {Ni}}, \bibinfo {author} {\bibfnamefont {Y.}~\bibnamefont
  {Chen}}, \bibinfo {author} {\bibfnamefont {C.-W.}\ \bibnamefont {Qiu}},\ and\
  \bibinfo {author} {\bibfnamefont {J.}~\bibnamefont {Rho}},\ }\bibfield
  {title} {\bibinfo {title} {{Electromagnetic chirality: from fundamentals to
  nontraditional chiroptical phenomena}},\ }\href
  {https://doi.org/10.1038/s41377-020-00367-8} {\bibfield  {journal} {\bibinfo
  {journal} {Light Sci. Appl.}\ }\textbf {\bibinfo {volume} {9}},\ \bibinfo
  {pages} {1} (\bibinfo {year} {2020})}\BibitemShut {NoStop}%
\bibitem [{\citenamefont {Bliokh}\ \emph {et~al.}(2014)\citenamefont {Bliokh},
  \citenamefont {Bekshaev},\ and\ \citenamefont
  {Nori}}]{bliokh2014ExtraordinaryMomentumSpin}%
  \BibitemOpen
  \bibfield  {author} {\bibinfo {author} {\bibfnamefont {K.~Y.}\ \bibnamefont
  {Bliokh}}, \bibinfo {author} {\bibfnamefont {A.~Y.}\ \bibnamefont
  {Bekshaev}},\ and\ \bibinfo {author} {\bibfnamefont {F.}~\bibnamefont
  {Nori}},\ }\bibfield  {title} {\bibinfo {title} {Extraordinary momentum and
  spin in evanescent waves},\ }\href {https://doi.org/10.1038/ncomms4300}
  {\bibfield  {journal} {\bibinfo  {journal} {Nature Communications}\ }\textbf
  {\bibinfo {volume} {5}},\ \bibinfo {pages} {1} (\bibinfo {year} {2014})},\
  \Eprint {https://arxiv.org/abs/1308.0547} {arXiv:1308.0547} \BibitemShut
  {NoStop}%
\bibitem [{\citenamefont {Li}\ \emph {et~al.}(2014{\natexlab{a}})\citenamefont
  {Li}, \citenamefont {Wu}, \citenamefont {Qu}, \citenamefont {Li},
  \citenamefont {Bai},\ and\ \citenamefont {Gong}}]{Li2014Dec}%
  \BibitemOpen
  \bibfield  {author} {\bibinfo {author} {\bibfnamefont {Z.}~\bibnamefont
  {Li}}, \bibinfo {author} {\bibfnamefont {Z.}~\bibnamefont {Wu}}, \bibinfo
  {author} {\bibfnamefont {T.}~\bibnamefont {Qu}}, \bibinfo {author}
  {\bibfnamefont {H.}~\bibnamefont {Li}}, \bibinfo {author} {\bibfnamefont
  {L.}~\bibnamefont {Bai}},\ and\ \bibinfo {author} {\bibfnamefont
  {L.}~\bibnamefont {Gong}},\ }\bibfield  {title} {\bibinfo {title} {{Radiation
  torque exerted on a uniaxial anisotropic sphere: Effects of various
  parameters}},\ }\href {https://doi.org/10.1016/j.optlastec.2014.05.026}
  {\bibfield  {journal} {\bibinfo  {journal} {Opt. Laser Technol.}\ }\textbf
  {\bibinfo {volume} {64}},\ \bibinfo {pages} {269} (\bibinfo {year}
  {2014}{\natexlab{a}})}\BibitemShut {NoStop}%
\bibitem [{\citenamefont {Li}\ \emph {et~al.}(2012{\natexlab{a}})\citenamefont
  {Li}, \citenamefont {Wu}, \citenamefont {Shang}, \citenamefont {Bai},\ and\
  \citenamefont {Cao}}]{Li2012Jul}%
  \BibitemOpen
  \bibfield  {author} {\bibinfo {author} {\bibfnamefont {Z.-J.}\ \bibnamefont
  {Li}}, \bibinfo {author} {\bibfnamefont {Z.-S.}\ \bibnamefont {Wu}}, \bibinfo
  {author} {\bibfnamefont {Q.-C.}\ \bibnamefont {Shang}}, \bibinfo {author}
  {\bibfnamefont {L.}~\bibnamefont {Bai}},\ and\ \bibinfo {author}
  {\bibfnamefont {C.-H.}\ \bibnamefont {Cao}},\ }\bibfield  {title} {\bibinfo
  {title} {{Calculation of radiation force and torque exerted on a uniaxial
  anisotropic sphere by an incident Gaussian beam with arbitrary propagation
  and polarization directions}},\ }\href {https://doi.org/10.1364/OE.20.016421}
  {\bibfield  {journal} {\bibinfo  {journal} {Opt. Express}\ }\textbf {\bibinfo
  {volume} {20}},\ \bibinfo {pages} {16421} (\bibinfo {year}
  {2012}{\natexlab{a}})}\BibitemShut {NoStop}%
\bibitem [{\citenamefont {Pesce}\ \emph {et~al.}(2020)\citenamefont {Pesce},
  \citenamefont {Jones}, \citenamefont
  {Marag{\ifmmode\grave{o}\else\`{o}\fi}},\ and\ \citenamefont
  {Volpe}}]{Pesce2020Dec}%
  \BibitemOpen
  \bibfield  {author} {\bibinfo {author} {\bibfnamefont {G.}~\bibnamefont
  {Pesce}}, \bibinfo {author} {\bibfnamefont {P.~H.}\ \bibnamefont {Jones}},
  \bibinfo {author} {\bibfnamefont {O.~M.}\ \bibnamefont
  {Marag{\ifmmode\grave{o}\else\`{o}\fi}}},\ and\ \bibinfo {author}
  {\bibfnamefont {G.}~\bibnamefont {Volpe}},\ }\bibfield  {title} {\bibinfo
  {title} {{Optical tweezers: theory and practice}},\ }\href
  {https://doi.org/10.1140/epjp/s13360-020-00843-5} {\bibfield  {journal}
  {\bibinfo  {journal} {Eur. Phys. J. Plus}\ }\textbf {\bibinfo {volume}
  {135}},\ \bibinfo {pages} {1} (\bibinfo {year} {2020})}\BibitemShut {NoStop}%
\bibitem [{\citenamefont {Borghese}\ \emph {et~al.}(2006)\citenamefont
  {Borghese}, \citenamefont {Denti}, \citenamefont {Saija},\ and\ \citenamefont
  {Iat{\ifmmode\grave{\imath}\else\`{\i}\fi}}}]{Borghese2006Oct}%
  \BibitemOpen
  \bibfield  {author} {\bibinfo {author} {\bibfnamefont {F.}~\bibnamefont
  {Borghese}}, \bibinfo {author} {\bibfnamefont {P.}~\bibnamefont {Denti}},
  \bibinfo {author} {\bibfnamefont {R.}~\bibnamefont {Saija}},\ and\ \bibinfo
  {author} {\bibfnamefont {M.~A.}\ \bibnamefont
  {Iat{\ifmmode\grave{\imath}\else\`{\i}\fi}}},\ }\bibfield  {title} {\bibinfo
  {title} {{Radiation torque on nonspherical particles in the transition matrix
  formalism}},\ }\href {https://doi.org/10.1364/OE.14.009508} {\bibfield
  {journal} {\bibinfo  {journal} {Opt. Express}\ }\textbf {\bibinfo {volume}
  {14}},\ \bibinfo {pages} {9508} (\bibinfo {year} {2006})}\BibitemShut
  {NoStop}%
\bibitem [{\citenamefont {Tsimokha}\ \emph {et~al.}(2022)\citenamefont
  {Tsimokha}, \citenamefont {Igoshin}, \citenamefont {Nikitina}, \citenamefont
  {Toftul}, \citenamefont {Frizyuk},\ and\ \citenamefont
  {Petrov}}]{Tsimokha2022Apr}%
  \BibitemOpen
  \bibfield  {author} {\bibinfo {author} {\bibfnamefont {M.}~\bibnamefont
  {Tsimokha}}, \bibinfo {author} {\bibfnamefont {V.}~\bibnamefont {Igoshin}},
  \bibinfo {author} {\bibfnamefont {A.}~\bibnamefont {Nikitina}}, \bibinfo
  {author} {\bibfnamefont {I.}~\bibnamefont {Toftul}}, \bibinfo {author}
  {\bibfnamefont {K.}~\bibnamefont {Frizyuk}},\ and\ \bibinfo {author}
  {\bibfnamefont {M.}~\bibnamefont {Petrov}},\ }\bibfield  {title} {\bibinfo
  {title} {{Acoustic resonators: Symmetry classification and multipolar content
  of the eigenmodes}},\ }\href {https://doi.org/10.1103/PhysRevB.105.165311}
  {\bibfield  {journal} {\bibinfo  {journal} {Phys. Rev. B}\ }\textbf {\bibinfo
  {volume} {105}},\ \bibinfo {pages} {165311} (\bibinfo {year}
  {2022})}\BibitemShut {NoStop}%
\bibitem [{\citenamefont {Sadrieva}\ \emph
  {et~al.}(2019{\natexlab{a}})\citenamefont {Sadrieva}, \citenamefont
  {Frizyuk}, \citenamefont {Petrov}, \citenamefont {Kivshar},\ and\
  \citenamefont {Bogdanov}}]{Sadrieva2019Sep}%
  \BibitemOpen
  \bibfield  {author} {\bibinfo {author} {\bibfnamefont {Z.}~\bibnamefont
  {Sadrieva}}, \bibinfo {author} {\bibfnamefont {K.}~\bibnamefont {Frizyuk}},
  \bibinfo {author} {\bibfnamefont {M.}~\bibnamefont {Petrov}}, \bibinfo
  {author} {\bibfnamefont {Y.}~\bibnamefont {Kivshar}},\ and\ \bibinfo {author}
  {\bibfnamefont {A.}~\bibnamefont {Bogdanov}},\ }\bibfield  {title} {\bibinfo
  {title} {{Multipolar origin of bound states in the continuum}},\ }\href
  {https://doi.org/10.1103/PhysRevB.100.115303} {\bibfield  {journal} {\bibinfo
   {journal} {Phys. Rev. B}\ }\textbf {\bibinfo {volume} {100}},\ \bibinfo
  {pages} {115303} (\bibinfo {year} {2019}{\natexlab{a}})}\BibitemShut
  {NoStop}%
\bibitem [{\citenamefont {Frizyuk}(2019)}]{Frizyuk2019Aug}%
  \BibitemOpen
  \bibfield  {author} {\bibinfo {author} {\bibfnamefont {K.}~\bibnamefont
  {Frizyuk}},\ }\bibfield  {title} {\bibinfo {title} {{Second-harmonic
  generation in dielectric nanoparticles with different symmetries}},\ }\href
  {https://doi.org/10.1364/JOSAB.36.000F32} {\bibfield  {journal} {\bibinfo
  {journal} {J. Opt. Soc. Am. B, JOSAB}\ }\textbf {\bibinfo {volume} {36}},\
  \bibinfo {pages} {F32} (\bibinfo {year} {2019})}\BibitemShut {NoStop}%
\bibitem [{\citenamefont {Frizyuk}\ \emph
  {et~al.}(2021{\natexlab{a}})\citenamefont {Frizyuk}, \citenamefont
  {Melik-Gaykazyan}, \citenamefont {Choi}, \citenamefont {Petrov},
  \citenamefont {Park},\ and\ \citenamefont {Kivshar}}]{Frizyuk2021May}%
  \BibitemOpen
  \bibfield  {author} {\bibinfo {author} {\bibfnamefont {K.}~\bibnamefont
  {Frizyuk}}, \bibinfo {author} {\bibfnamefont {E.}~\bibnamefont
  {Melik-Gaykazyan}}, \bibinfo {author} {\bibfnamefont {J.-H.}\ \bibnamefont
  {Choi}}, \bibinfo {author} {\bibfnamefont {M.~I.}\ \bibnamefont {Petrov}},
  \bibinfo {author} {\bibfnamefont {H.-G.}\ \bibnamefont {Park}},\ and\
  \bibinfo {author} {\bibfnamefont {Y.}~\bibnamefont {Kivshar}},\ }\bibfield
  {title} {\bibinfo {title} {{Nonlinear Circular Dichroism in Mie-Resonant
  Nanoparticle Dimers}},\ }\href {https://doi.org/10.1021/acs.nanolett.1c01025}
  {\bibfield  {journal} {\bibinfo  {journal} {Nano Lett.}\ }\textbf {\bibinfo
  {volume} {21}},\ \bibinfo {pages} {4381} (\bibinfo {year}
  {2021}{\natexlab{a}})}\BibitemShut {NoStop}%
\bibitem [{\citenamefont {Finazzi}\ \emph {et~al.}(2007)\citenamefont
  {Finazzi}, \citenamefont {Biagioni}, \citenamefont {Celebrano},\ and\
  \citenamefont {Du{\ifmmode\grave{o}\else\`{o}\fi}}}]{Finazzi2007Sep}%
  \BibitemOpen
  \bibfield  {author} {\bibinfo {author} {\bibfnamefont {M.}~\bibnamefont
  {Finazzi}}, \bibinfo {author} {\bibfnamefont {P.}~\bibnamefont {Biagioni}},
  \bibinfo {author} {\bibfnamefont {M.}~\bibnamefont {Celebrano}},\ and\
  \bibinfo {author} {\bibfnamefont {L.}~\bibnamefont
  {Du{\ifmmode\grave{o}\else\`{o}\fi}}},\ }\bibfield  {title} {\bibinfo {title}
  {{Selection rules for second-harmonic generation in nanoparticles}},\ }\href
  {https://doi.org/10.1103/PhysRevB.76.125414} {\bibfield  {journal} {\bibinfo
  {journal} {Phys. Rev. B}\ }\textbf {\bibinfo {volume} {76}},\ \bibinfo
  {pages} {125414} (\bibinfo {year} {2007})}\BibitemShut {NoStop}%
\bibitem [{\citenamefont {Makarov}\ \emph {et~al.}(2017)\citenamefont
  {Makarov}, \citenamefont {Petrov}, \citenamefont {Zywietz}, \citenamefont
  {Milichko}, \citenamefont {Zuev}, \citenamefont {Lopanitsyna}, \citenamefont
  {Kuksin}, \citenamefont {Mukhin}, \citenamefont {Zograf}, \citenamefont
  {Ubyivovk}, \citenamefont {Smirnova}, \citenamefont {Starikov}, \citenamefont
  {Chichkov},\ and\ \citenamefont {Kivshar}}]{Makarov2017May}%
  \BibitemOpen
  \bibfield  {author} {\bibinfo {author} {\bibfnamefont {S.~V.}\ \bibnamefont
  {Makarov}}, \bibinfo {author} {\bibfnamefont {M.~I.}\ \bibnamefont {Petrov}},
  \bibinfo {author} {\bibfnamefont {U.}~\bibnamefont {Zywietz}}, \bibinfo
  {author} {\bibfnamefont {V.}~\bibnamefont {Milichko}}, \bibinfo {author}
  {\bibfnamefont {D.}~\bibnamefont {Zuev}}, \bibinfo {author} {\bibfnamefont
  {N.}~\bibnamefont {Lopanitsyna}}, \bibinfo {author} {\bibfnamefont
  {A.}~\bibnamefont {Kuksin}}, \bibinfo {author} {\bibfnamefont
  {I.}~\bibnamefont {Mukhin}}, \bibinfo {author} {\bibfnamefont
  {G.}~\bibnamefont {Zograf}}, \bibinfo {author} {\bibfnamefont
  {E.}~\bibnamefont {Ubyivovk}}, \bibinfo {author} {\bibfnamefont {D.~A.}\
  \bibnamefont {Smirnova}}, \bibinfo {author} {\bibfnamefont {S.}~\bibnamefont
  {Starikov}}, \bibinfo {author} {\bibfnamefont {B.~N.}\ \bibnamefont
  {Chichkov}},\ and\ \bibinfo {author} {\bibfnamefont {Y.~S.}\ \bibnamefont
  {Kivshar}},\ }\bibfield  {title} {\bibinfo {title} {{Efficient
  Second-Harmonic Generation in Nanocrystalline Silicon Nanoparticles}},\
  }\href {https://doi.org/10.1021/acs.nanolett.7b00392} {\bibfield  {journal}
  {\bibinfo  {journal} {Nano Lett.}\ }\textbf {\bibinfo {volume} {17}},\
  \bibinfo {pages} {3047} (\bibinfo {year} {2017})}\BibitemShut {NoStop}%
\bibitem [{\citenamefont {Smirnova}\ \emph {et~al.}(2018)\citenamefont
  {Smirnova}, \citenamefont {Smirnov},\ and\ \citenamefont
  {Kivshar}}]{Smirnova2018Jan}%
  \BibitemOpen
  \bibfield  {author} {\bibinfo {author} {\bibfnamefont {D.}~\bibnamefont
  {Smirnova}}, \bibinfo {author} {\bibfnamefont {A.~I.}\ \bibnamefont
  {Smirnov}},\ and\ \bibinfo {author} {\bibfnamefont {Y.~S.}\ \bibnamefont
  {Kivshar}},\ }\bibfield  {title} {\bibinfo {title} {{Multipolar
  second-harmonic generation by Mie-resonant dielectric nanoparticles}},\
  }\href {https://doi.org/10.1103/PhysRevA.97.013807} {\bibfield  {journal}
  {\bibinfo  {journal} {Phys. Rev. A}\ }\textbf {\bibinfo {volume} {97}},\
  \bibinfo {pages} {013807} (\bibinfo {year} {2018})}\BibitemShut {NoStop}%
\bibitem [{\citenamefont {Smith}(1958)}]{Smith1958-MacroscopicSymmetry}%
  \BibitemOpen
  \bibfield  {author} {\bibinfo {author} {\bibfnamefont {C.~S.}\ \bibnamefont
  {Smith}},\ }\bibfield  {title} {\bibinfo {title} {{Macroscopic Symmetry and
  Properties of Crystals}},\ }in\ \href
  {https://doi.org/10.1016/S0081-1947(08)60727-4} {\emph {\bibinfo {booktitle}
  {{Solid State Physics}}}},\ Vol.~\bibinfo {volume} {6}\ (\bibinfo
  {publisher} {Academic Press},\ \bibinfo {address} {Cambridge, MA, USA},\
  \bibinfo {year} {1958})\ pp.\ \bibinfo {pages} {175--249}\BibitemShut
  {NoStop}%
\bibitem [{\citenamefont {Boyd}(2020)}]{boyd2020nonlinear}%
  \BibitemOpen
  \bibfield  {author} {\bibinfo {author} {\bibfnamefont {R.}~\bibnamefont
  {Boyd}},\ }\href {https://books.google.com.au/books?id=54vZDwAAQBAJ} {\emph
  {\bibinfo {title} {Nonlinear Optics}}}\ (\bibinfo  {publisher} {Elsevier
  Science},\ \bibinfo {year} {2020})\BibitemShut {NoStop}%
\bibitem [{\citenamefont {Miller}(1964)}]{Miller1964-OPTICALSECONDHARMON}%
  \BibitemOpen
  \bibfield  {author} {\bibinfo {author} {\bibfnamefont {R.~C.}\ \bibnamefont
  {Miller}},\ }\bibfield  {title} {\bibinfo {title} {{Optical second harmonic
  generation in piezoelectric crystals}},\ }\href
  {https://doi.org/10.1063/1.1754022} {\bibfield  {journal} {\bibinfo
  {journal} {Appl. Phys. Lett.}\ }\textbf {\bibinfo {volume} {5}},\ \bibinfo
  {pages} {17} (\bibinfo {year} {1964})}\BibitemShut {NoStop}%
\bibitem [{\citenamefont {Franken}\ and\ \citenamefont
  {Ward}(1963)}]{Franken1963-OpticalHarmonicsand}%
  \BibitemOpen
  \bibfield  {author} {\bibinfo {author} {\bibfnamefont {P.~A.}\ \bibnamefont
  {Franken}}\ and\ \bibinfo {author} {\bibfnamefont {J.~F.}\ \bibnamefont
  {Ward}},\ }\bibfield  {title} {\bibinfo {title} {{Optical Harmonics and
  Nonlinear Phenomena}},\ }\href {https://doi.org/10.1103/RevModPhys.35.23}
  {\bibfield  {journal} {\bibinfo  {journal} {Rev. Mod. Phys.}\ }\textbf
  {\bibinfo {volume} {35}},\ \bibinfo {pages} {23} (\bibinfo {year}
  {1963})}\BibitemShut {NoStop}%
\bibitem [{\citenamefont {Frizyuk}\ \emph
  {et~al.}(2021{\natexlab{b}})\citenamefont {Frizyuk}, \citenamefont
  {Melik-Gaykazyan}, \citenamefont {Choi}, \citenamefont {Petrov},
  \citenamefont {Park},\ and\ \citenamefont
  {Kivshar}}]{Frizyuk2021-NonlinearCircularDi}%
  \BibitemOpen
  \bibfield  {author} {\bibinfo {author} {\bibfnamefont {K.}~\bibnamefont
  {Frizyuk}}, \bibinfo {author} {\bibfnamefont {E.}~\bibnamefont
  {Melik-Gaykazyan}}, \bibinfo {author} {\bibfnamefont {J.-H.}\ \bibnamefont
  {Choi}}, \bibinfo {author} {\bibfnamefont {M.~I.}\ \bibnamefont {Petrov}},
  \bibinfo {author} {\bibfnamefont {H.-G.}\ \bibnamefont {Park}},\ and\
  \bibinfo {author} {\bibfnamefont {Y.}~\bibnamefont {Kivshar}},\ }\bibfield
  {title} {\bibinfo {title} {{Nonlinear Circular Dichroism in Mie-Resonant
  Nanoparticle Dimers}},\ }\href {https://doi.org/10.1021/acs.nanolett.1c01025}
  {\bibfield  {journal} {\bibinfo  {journal} {Nano Lett.}\ }\textbf {\bibinfo
  {volume} {21}},\ \bibinfo {pages} {4381} (\bibinfo {year}
  {2021}{\natexlab{b}})}\BibitemShut {NoStop}%
\bibitem [{\citenamefont {Gladyshev}\ \emph {et~al.}(2020)\citenamefont
  {Gladyshev}, \citenamefont {Frizyuk},\ and\ \citenamefont
  {Bogdanov}}]{Gladyshev2020}%
  \BibitemOpen
  \bibfield  {author} {\bibinfo {author} {\bibfnamefont {S.}~\bibnamefont
  {Gladyshev}}, \bibinfo {author} {\bibfnamefont {K.}~\bibnamefont {Frizyuk}},\
  and\ \bibinfo {author} {\bibfnamefont {A.}~\bibnamefont {Bogdanov}},\
  }\bibfield  {title} {\bibinfo {title} {{Symmetry analysis and multipole
  classification of eigenmodes in electromagnetic resonators for engineering
  their optical properties}},\ }\href
  {https://doi.org/10.1103/PhysRevB.102.075103} {\bibfield  {journal} {\bibinfo
   {journal} {Physical Review B}\ }\textbf {\bibinfo {volume} {102}},\ \bibinfo
  {pages} {75103} (\bibinfo {year} {2020})}\BibitemShut {NoStop}%
\bibitem [{\citenamefont {Sadrieva}\ \emph
  {et~al.}(2019{\natexlab{b}})\citenamefont {Sadrieva}, \citenamefont
  {Frizyuk}, \citenamefont {Petrov}, \citenamefont {Kivshar},\ and\
  \citenamefont {Bogdanov}}]{Sadrieva2019}%
  \BibitemOpen
  \bibfield  {author} {\bibinfo {author} {\bibfnamefont {Z.}~\bibnamefont
  {Sadrieva}}, \bibinfo {author} {\bibfnamefont {K.}~\bibnamefont {Frizyuk}},
  \bibinfo {author} {\bibfnamefont {M.}~\bibnamefont {Petrov}}, \bibinfo
  {author} {\bibfnamefont {Y.}~\bibnamefont {Kivshar}},\ and\ \bibinfo {author}
  {\bibfnamefont {A.}~\bibnamefont {Bogdanov}},\ }\bibfield  {title} {\bibinfo
  {title} {{Multipolar origin of bound states in the continuum}},\ }\href
  {https://doi.org/10.1103/PhysRevB.100.115303} {\bibfield  {journal} {\bibinfo
   {journal} {Physical Review B}\ }\textbf {\bibinfo {volume} {100}},\ \bibinfo
  {pages} {1} (\bibinfo {year} {2019}{\natexlab{b}})},\ \Eprint
  {https://arxiv.org/abs/1903.00309} {arXiv:1903.00309} \BibitemShut {NoStop}%
\bibitem [{\citenamefont {Liu}\ \emph {et~al.}(2020)\citenamefont {Liu},
  \citenamefont {Xu}, \citenamefont {Yu}, \citenamefont {Wang},\ and\
  \citenamefont {Takahara}}]{Liu2020}%
  \BibitemOpen
  \bibfield  {author} {\bibinfo {author} {\bibfnamefont {T.}~\bibnamefont
  {Liu}}, \bibinfo {author} {\bibfnamefont {R.}~\bibnamefont {Xu}}, \bibinfo
  {author} {\bibfnamefont {P.}~\bibnamefont {Yu}}, \bibinfo {author}
  {\bibfnamefont {Z.}~\bibnamefont {Wang}},\ and\ \bibinfo {author}
  {\bibfnamefont {J.}~\bibnamefont {Takahara}},\ }\bibfield  {title} {\bibinfo
  {title} {{Multipole and multimode engineering in Mie resonance-based
  metastructures}},\ }\href {https://doi.org/10.1515/nanoph-2019-0505}
  {\bibfield  {journal} {\bibinfo  {journal} {Nanophotonics}\ }\textbf
  {\bibinfo {volume} {0}},\ \bibinfo {pages} {1} (\bibinfo {year}
  {2020})}\BibitemShut {NoStop}%
\bibitem [{\citenamefont {Koshelev}\ \emph {et~al.}(2022)\citenamefont
  {Koshelev}, \citenamefont {Sadrieva}, \citenamefont {Shcherbakov},
  \citenamefont {Kivshar},\ and\ \citenamefont {Bogdanov}}]{Koshelev2022Jul}%
  \BibitemOpen
  \bibfield  {author} {\bibinfo {author} {\bibfnamefont {K.}~\bibnamefont
  {Koshelev}}, \bibinfo {author} {\bibfnamefont {Z.}~\bibnamefont {Sadrieva}},
  \bibinfo {author} {\bibfnamefont {A.}~\bibnamefont {Shcherbakov}}, \bibinfo
  {author} {\bibfnamefont {Y.}~\bibnamefont {Kivshar}},\ and\ \bibinfo {author}
  {\bibfnamefont {A.}~\bibnamefont {Bogdanov}},\ }\bibfield  {title} {\bibinfo
  {title} {{Bound states in the continuum in photonic structures}},\ }\bibfield
   {journal} {\bibinfo  {journal} {arXiv}\ }\href
  {https://doi.org/10.3367/UFNe.2021.12.039120} {10.3367/UFNe.2021.12.039120}
  (\bibinfo {year} {2022}),\ \Eprint {https://arxiv.org/abs/2207.01441}
  {2207.01441} \BibitemShut {NoStop}%
\bibitem [{\citenamefont {Rybin}\ \emph {et~al.}(2017)\citenamefont {Rybin},
  \citenamefont {Koshelev}, \citenamefont {Sadrieva}, \citenamefont {Samusev},
  \citenamefont {Bogdanov}, \citenamefont {Limonov},\ and\ \citenamefont
  {Kivshar}}]{Rybin2017Dec}%
  \BibitemOpen
  \bibfield  {author} {\bibinfo {author} {\bibfnamefont {M.~V.}\ \bibnamefont
  {Rybin}}, \bibinfo {author} {\bibfnamefont {K.~L.}\ \bibnamefont {Koshelev}},
  \bibinfo {author} {\bibfnamefont {Z.~F.}\ \bibnamefont {Sadrieva}}, \bibinfo
  {author} {\bibfnamefont {K.~B.}\ \bibnamefont {Samusev}}, \bibinfo {author}
  {\bibfnamefont {A.~A.}\ \bibnamefont {Bogdanov}}, \bibinfo {author}
  {\bibfnamefont {M.~F.}\ \bibnamefont {Limonov}},\ and\ \bibinfo {author}
  {\bibfnamefont {Y.~S.}\ \bibnamefont {Kivshar}},\ }\bibfield  {title}
  {\bibinfo {title} {{High-$Q$ Supercavity Modes in Subwavelength Dielectric
  Resonators}},\ }\href {https://doi.org/10.1103/PhysRevLett.119.243901}
  {\bibfield  {journal} {\bibinfo  {journal} {Phys. Rev. Lett.}\ }\textbf
  {\bibinfo {volume} {119}},\ \bibinfo {pages} {243901} (\bibinfo {year}
  {2017})}\BibitemShut {NoStop}%
\bibitem [{\citenamefont {Koshelev}\ \emph {et~al.}(2020)\citenamefont
  {Koshelev}, \citenamefont {Kruk}, \citenamefont {Melik-Gaykazyan},
  \citenamefont {Choi}, \citenamefont {Bogdanov}, \citenamefont {Park},\ and\
  \citenamefont {Kivshar}}]{Koshelev2020Jan}%
  \BibitemOpen
  \bibfield  {author} {\bibinfo {author} {\bibfnamefont {K.}~\bibnamefont
  {Koshelev}}, \bibinfo {author} {\bibfnamefont {S.}~\bibnamefont {Kruk}},
  \bibinfo {author} {\bibfnamefont {E.}~\bibnamefont {Melik-Gaykazyan}},
  \bibinfo {author} {\bibfnamefont {J.-H.}\ \bibnamefont {Choi}}, \bibinfo
  {author} {\bibfnamefont {A.}~\bibnamefont {Bogdanov}}, \bibinfo {author}
  {\bibfnamefont {H.-G.}\ \bibnamefont {Park}},\ and\ \bibinfo {author}
  {\bibfnamefont {Y.}~\bibnamefont {Kivshar}},\ }\bibfield  {title} {\bibinfo
  {title} {{Subwavelength dielectric resonators for nonlinear nanophotonics}},\
  }\href {https://doi.org/10.1126/science.aaz3985} {\bibfield  {journal}
  {\bibinfo  {journal} {Science}\ }\textbf {\bibinfo {volume} {367}},\ \bibinfo
  {pages} {288} (\bibinfo {year} {2020})}\BibitemShut {NoStop}%
\bibitem [{\citenamefont {Bogdanov}\ \emph {et~al.}(2019)\citenamefont
  {Bogdanov}, \citenamefont {Koshelev}, \citenamefont {Kapitanova},
  \citenamefont {Rybin}, \citenamefont {Gladyshev}, \citenamefont {Sadrieva},
  \citenamefont {Samusev}, \citenamefont {Kivshar},\ and\ \citenamefont
  {Limonov}}]{Bogdanov2019Jan}%
  \BibitemOpen
  \bibfield  {author} {\bibinfo {author} {\bibfnamefont {A.~A.}\ \bibnamefont
  {Bogdanov}}, \bibinfo {author} {\bibfnamefont {K.~L.}\ \bibnamefont
  {Koshelev}}, \bibinfo {author} {\bibfnamefont {P.~V.}\ \bibnamefont
  {Kapitanova}}, \bibinfo {author} {\bibfnamefont {M.~V.}\ \bibnamefont
  {Rybin}}, \bibinfo {author} {\bibfnamefont {S.~A.}\ \bibnamefont
  {Gladyshev}}, \bibinfo {author} {\bibfnamefont {Z.~F.}\ \bibnamefont
  {Sadrieva}}, \bibinfo {author} {\bibfnamefont {K.~B.}\ \bibnamefont
  {Samusev}}, \bibinfo {author} {\bibfnamefont {Y.~S.}\ \bibnamefont
  {Kivshar}},\ and\ \bibinfo {author} {\bibfnamefont {M.~F.}\ \bibnamefont
  {Limonov}},\ }\bibfield  {title} {\bibinfo {title} {{Bound states in the
  continuum and Fano resonances in the strong mode coupling regime}},\
  }\bibfield  {booktitle} {\emph {\bibinfo {booktitle} {{Advanced Photonics,
  1(1)}}},\ }\href {https://doi.org/10.1117/1.AP.1.1.016001} {\bibfield
  {journal} {\bibinfo  {journal} {Adv. Photonics}\ }\textbf {\bibinfo {volume}
  {1}},\ \bibinfo {pages} {016001} (\bibinfo {year} {2019})}\BibitemShut
  {NoStop}%
\bibitem [{\citenamefont {{Friedrich H.}}\ and\ \citenamefont {{Wintgen
  D.}}(1985)}]{FriedrichH.1985}%
  \BibitemOpen
  \bibfield  {author} {\bibinfo {author} {\bibnamefont {{Friedrich H.}}}\ and\
  \bibinfo {author} {\bibnamefont {{Wintgen D.}}},\ }\bibfield  {title}
  {\bibinfo {title} {{Interfering resonances and BIC}},\ }\href
  {https://journals.aps.org/pra/pdf/10.1103/PhysRevA.32.3231} {\bibfield
  {journal} {\bibinfo  {journal} {Physical Review A}\ }\textbf {\bibinfo
  {volume} {32}},\ \bibinfo {pages} {3231} (\bibinfo {year}
  {1985})}\BibitemShut {NoStop}%
\bibitem [{\citenamefont {McDonnell}\ and\ \citenamefont
  {Wallace}(2016)}]{McDonnell2016Oct}%
  \BibitemOpen
  \bibfield  {author} {\bibinfo {author} {\bibfnamefont {S.~J.}\ \bibnamefont
  {McDonnell}}\ and\ \bibinfo {author} {\bibfnamefont {R.~M.}\ \bibnamefont
  {Wallace}},\ }\bibfield  {title} {\bibinfo {title} {{Atomically-thin layered
  films for device applications based upon 2D TMDC materials}},\ }\href
  {https://doi.org/10.1016/j.tsf.2016.08.068} {\bibfield  {journal} {\bibinfo
  {journal} {Thin Solid Films}\ }\textbf {\bibinfo {volume} {616}},\ \bibinfo
  {pages} {482} (\bibinfo {year} {2016})}\BibitemShut {NoStop}%
\bibitem [{\citenamefont {Zhao}\ \emph {et~al.}(2020)\citenamefont {Zhao},
  \citenamefont {Chen}, \citenamefont {Xu}, \citenamefont {Zhang},
  \citenamefont {Shi}, \citenamefont {Shao},\ and\ \citenamefont
  {Wan}}]{Zhao2020Nov}%
  \BibitemOpen
  \bibfield  {author} {\bibinfo {author} {\bibfnamefont {X.-k.}\ \bibnamefont
  {Zhao}}, \bibinfo {author} {\bibfnamefont {R.-w.}\ \bibnamefont {Chen}},
  \bibinfo {author} {\bibfnamefont {K.}~\bibnamefont {Xu}}, \bibinfo {author}
  {\bibfnamefont {S.-y.}\ \bibnamefont {Zhang}}, \bibinfo {author}
  {\bibfnamefont {H.}~\bibnamefont {Shi}}, \bibinfo {author} {\bibfnamefont
  {Z.-y.}\ \bibnamefont {Shao}},\ and\ \bibinfo {author} {\bibfnamefont
  {N.}~\bibnamefont {Wan}},\ }\bibfield  {title} {\bibinfo {title} {{Nanoscale
  water film at a super-wetting interface supports 2D material transfer}},\
  }\href {https://doi.org/10.1088/2053-1583/abc2a7} {\bibfield  {journal}
  {\bibinfo  {journal} {2D Mater.}\ }\textbf {\bibinfo {volume} {8}},\ \bibinfo
  {pages} {015021} (\bibinfo {year} {2020})}\BibitemShut {NoStop}%
\bibitem [{\citenamefont {Jin}\ \emph {et~al.}(2019)\citenamefont {Jin},
  \citenamefont {Lee}, \citenamefont {Liao}, \citenamefont {Kim}, \citenamefont
  {Wang}, \citenamefont {Okello}, \citenamefont {Park}, \citenamefont {Park},
  \citenamefont {Han}, \citenamefont {Heo}, \citenamefont {Kim}, \citenamefont
  {Oh}, \citenamefont {Choi}, \citenamefont {Park},\ and\ \citenamefont
  {Jo}}]{Jin2019Jul}%
  \BibitemOpen
  \bibfield  {author} {\bibinfo {author} {\bibfnamefont {G.}~\bibnamefont
  {Jin}}, \bibinfo {author} {\bibfnamefont {C.-S.}\ \bibnamefont {Lee}},
  \bibinfo {author} {\bibfnamefont {X.}~\bibnamefont {Liao}}, \bibinfo {author}
  {\bibfnamefont {J.}~\bibnamefont {Kim}}, \bibinfo {author} {\bibfnamefont
  {Z.}~\bibnamefont {Wang}}, \bibinfo {author} {\bibfnamefont {O.~F.~N.}\
  \bibnamefont {Okello}}, \bibinfo {author} {\bibfnamefont {B.}~\bibnamefont
  {Park}}, \bibinfo {author} {\bibfnamefont {J.}~\bibnamefont {Park}}, \bibinfo
  {author} {\bibfnamefont {C.}~\bibnamefont {Han}}, \bibinfo {author}
  {\bibfnamefont {H.}~\bibnamefont {Heo}}, \bibinfo {author} {\bibfnamefont
  {J.}~\bibnamefont {Kim}}, \bibinfo {author} {\bibfnamefont {S.~H.}\
  \bibnamefont {Oh}}, \bibinfo {author} {\bibfnamefont {S.-Y.}\ \bibnamefont
  {Choi}}, \bibinfo {author} {\bibfnamefont {H.}~\bibnamefont {Park}},\ and\
  \bibinfo {author} {\bibfnamefont {M.-H.}\ \bibnamefont {Jo}},\ }\bibfield
  {title} {\bibinfo {title} {{Atomically thin three-dimensional membranes of
  van der Waals semiconductors by wafer-scale growth}},\ }\href
  {https://doi.org/10.1126/sciadv.aaw3180} {\bibfield  {journal} {\bibinfo
  {journal} {Sci. Adv.}\ }\textbf {\bibinfo {volume} {5}},\ \bibinfo {pages}
  {eaaw3180} (\bibinfo {year} {2019})}\BibitemShut {NoStop}%
\bibitem [{\citenamefont {You}\ \emph {et~al.}(2019)\citenamefont {You},
  \citenamefont {Bongu}, \citenamefont {Bao},\ and\ \citenamefont
  {Panoiu}}]{You2019Jan}%
  \BibitemOpen
  \bibfield  {author} {\bibinfo {author} {\bibfnamefont {J.~W.}\ \bibnamefont
  {You}}, \bibinfo {author} {\bibfnamefont {S.~R.}\ \bibnamefont {Bongu}},
  \bibinfo {author} {\bibfnamefont {Q.}~\bibnamefont {Bao}},\ and\ \bibinfo
  {author} {\bibfnamefont {N.~C.}\ \bibnamefont {Panoiu}},\ }\bibfield  {title}
  {\bibinfo {title} {{Nonlinear optical properties and applications of 2D
  materials: theoretical and experimental aspects}},\ }\href
  {https://doi.org/10.1515/nanoph-2018-0106} {\bibfield  {journal} {\bibinfo
  {journal} {Nanophotonics}\ }\textbf {\bibinfo {volume} {8}},\ \bibinfo
  {pages} {63} (\bibinfo {year} {2019})}\BibitemShut {NoStop}%
\bibitem [{\citenamefont {Ermolaev}\ \emph {et~al.}(2019)\citenamefont
  {Ermolaev}, \citenamefont {Yakubovsky}, \citenamefont {Stebunov},
  \citenamefont {Arsenin},\ and\ \citenamefont {Volkov}}]{Ermolaev2019Dec}%
  \BibitemOpen
  \bibfield  {author} {\bibinfo {author} {\bibfnamefont {G.~A.}\ \bibnamefont
  {Ermolaev}}, \bibinfo {author} {\bibfnamefont {D.~I.}\ \bibnamefont
  {Yakubovsky}}, \bibinfo {author} {\bibfnamefont {Y.~V.}\ \bibnamefont
  {Stebunov}}, \bibinfo {author} {\bibfnamefont {A.~V.}\ \bibnamefont
  {Arsenin}},\ and\ \bibinfo {author} {\bibfnamefont {V.~S.}\ \bibnamefont
  {Volkov}},\ }\bibfield  {title} {\bibinfo {title} {{Spectral ellipsometry of
  monolayer transition metal dichalcogenides: Analysis of excitonic peaks in
  dispersion}},\ }\href {https://doi.org/10.1116/1.5122683} {\bibfield
  {journal} {\bibinfo  {journal} {Journal of Vacuum Science {\&} Technology B,
  Nanotechnology and Microelectronics: Materials, Processing, Measurement, and
  Phenomena}\ }\textbf {\bibinfo {volume} {38}},\ \bibinfo {pages} {014002}
  (\bibinfo {year} {2019})}\BibitemShut {NoStop}%
\bibitem [{\citenamefont {Nikitina}\ \emph {et~al.}(2022)\citenamefont
  {Nikitina}, \citenamefont {Nikolaeva},\ and\ \citenamefont
  {Frizyuk}}]{Nikitina2022-Whendoesnonlinearc}%
  \BibitemOpen
  \bibfield  {author} {\bibinfo {author} {\bibfnamefont {A.}~\bibnamefont
  {Nikitina}}, \bibinfo {author} {\bibfnamefont {A.}~\bibnamefont
  {Nikolaeva}},\ and\ \bibinfo {author} {\bibfnamefont {K.}~\bibnamefont
  {Frizyuk}},\ }\bibfield  {title} {\bibinfo {title} {{When does nonlinear
  circular dichroism appear in achiral dielectric nanoparticles?}},\ }\bibfield
   {journal} {\bibinfo  {journal} {arXiv}\ }\href
  {https://doi.org/10.48550/arXiv.2208.00891} {10.48550/arXiv.2208.00891}
  (\bibinfo {year} {2022}),\ \Eprint {https://arxiv.org/abs/2208.00891}
  {2208.00891} \BibitemShut {NoStop}%
\bibitem [{\citenamefont {Bohren}\ and\ \citenamefont
  {Huffman}(2008)}]{bohren2008absorption}%
  \BibitemOpen
  \bibfield  {author} {\bibinfo {author} {\bibfnamefont {C.~F.}\ \bibnamefont
  {Bohren}}\ and\ \bibinfo {author} {\bibfnamefont {D.~R.}\ \bibnamefont
  {Huffman}},\ }\href@noop {} {\emph {\bibinfo {title} {Absorption and
  scattering of light by small particles}}}\ (\bibinfo  {publisher} {John Wiley
  \& Sons},\ \bibinfo {year} {2008})\BibitemShut {NoStop}%
\bibitem [{\citenamefont {Doost}\ \emph {et~al.}(2014)\citenamefont {Doost},
  \citenamefont {Langbein},\ and\ \citenamefont {Muljarov}}]{Doost2014Jul}%
  \BibitemOpen
  \bibfield  {author} {\bibinfo {author} {\bibfnamefont {M.~B.}\ \bibnamefont
  {Doost}}, \bibinfo {author} {\bibfnamefont {W.}~\bibnamefont {Langbein}},\
  and\ \bibinfo {author} {\bibfnamefont {E.~A.}\ \bibnamefont {Muljarov}},\
  }\bibfield  {title} {\bibinfo {title} {{Resonant-state expansion applied to
  three-dimensional open optical systems}},\ }\href
  {https://doi.org/10.1103/PhysRevA.90.013834} {\bibfield  {journal} {\bibinfo
  {journal} {Phys. Rev. A}\ }\textbf {\bibinfo {volume} {90}},\ \bibinfo
  {pages} {013834} (\bibinfo {year} {2014})}\BibitemShut {NoStop}%
\bibitem [{\citenamefont {Sehmi}\ \emph {et~al.}(2020)\citenamefont {Sehmi},
  \citenamefont {Langbein},\ and\ \citenamefont {Muljarov}}]{Sehmi2020Jan}%
  \BibitemOpen
  \bibfield  {author} {\bibinfo {author} {\bibfnamefont {H.~S.}\ \bibnamefont
  {Sehmi}}, \bibinfo {author} {\bibfnamefont {W.}~\bibnamefont {Langbein}},\
  and\ \bibinfo {author} {\bibfnamefont {E.~A.}\ \bibnamefont {Muljarov}},\
  }\bibfield  {title} {\bibinfo {title} {{Applying the resonant-state expansion
  to realistic materials with frequency dispersion}},\ }\href
  {https://doi.org/10.1103/PhysRevB.101.045304} {\bibfield  {journal} {\bibinfo
   {journal} {Phys. Rev. B}\ }\textbf {\bibinfo {volume} {101}},\ \bibinfo
  {pages} {045304} (\bibinfo {year} {2020})}\BibitemShut {NoStop}%
\bibitem [{\citenamefont {Lobanov}\ \emph {et~al.}(2019)\citenamefont
  {Lobanov}, \citenamefont {Langbein},\ and\ \citenamefont
  {Muljarov}}]{Lobanov2019Dec}%
  \BibitemOpen
  \bibfield  {author} {\bibinfo {author} {\bibfnamefont {S.~V.}\ \bibnamefont
  {Lobanov}}, \bibinfo {author} {\bibfnamefont {W.}~\bibnamefont {Langbein}},\
  and\ \bibinfo {author} {\bibfnamefont {E.~A.}\ \bibnamefont {Muljarov}},\
  }\bibfield  {title} {\bibinfo {title} {{Resonant-state expansion applied to
  three-dimensional open optical systems: Complete set of static modes}},\
  }\href {https://doi.org/10.1103/PhysRevA.100.063811} {\bibfield  {journal}
  {\bibinfo  {journal} {Phys. Rev. A}\ }\textbf {\bibinfo {volume} {100}},\
  \bibinfo {pages} {063811} (\bibinfo {year} {2019})}\BibitemShut {NoStop}%
\bibitem [{\citenamefont {Varshalovich}\ \emph {et~al.}(1988)\citenamefont
  {Varshalovich}, \citenamefont {Moskalev},\ and\ \citenamefont
  {Khersonskii}}]{Varshalovich1988Oct}%
  \BibitemOpen
  \bibfield  {author} {\bibinfo {author} {\bibfnamefont {D.~A.}\ \bibnamefont
  {Varshalovich}}, \bibinfo {author} {\bibfnamefont {A.~N.}\ \bibnamefont
  {Moskalev}},\ and\ \bibinfo {author} {\bibfnamefont {V.~K.}\ \bibnamefont
  {Khersonskii}},\ }\href {https://doi.org/10.1142/0270} {\emph {\bibinfo
  {title} {{Quantum Theory of Angular Momentum}}}}\ (\bibinfo  {publisher}
  {World Scientific Publishing Company},\ \bibinfo {address} {Singapore},\
  \bibinfo {year} {1988})\BibitemShut {NoStop}%
\bibitem [{\citenamefont {Weisstein}(2003)}]{Weisstein2003Nov}%
  \BibitemOpen
  \bibfield  {author} {\bibinfo {author} {\bibfnamefont {E.~W.}\ \bibnamefont
  {Weisstein}},\ }\bibfield  {title} {\bibinfo {title} {{Condon-Shortley
  Phase}},\ }\href {https://mathworld.wolfram.com/Condon-ShortleyPhase.html}
  {\bibfield  {journal} {\bibinfo  {journal} {Wolfram Research, Inc.}\ }
  (\bibinfo {year} {2003})}\BibitemShut {NoStop}%
\bibitem [{\citenamefont {Stratton}()}]{stratton2007ElectromagneticTheory}%
  \BibitemOpen
  \bibfield  {author} {\bibinfo {author} {\bibfnamefont {J.~A.}\ \bibnamefont
  {Stratton}},\ }\href@noop {} {\emph {\bibinfo {title} {Electromagnetic
  Theory}}}\ (\bibinfo  {publisher} {{John Wiley \textbackslash\&
  Sons}})\BibitemShut {NoStop}%
\bibitem [{Bib(2021)}]{BibEntry2021Jun_comsol}%
  \BibitemOpen
  \href
  {https://doc.comsol.com/5.5/docserver/#!/com.comsol.help.comsol/comsol_ref_definitions.12.037.html}
  {\bibinfo {title} {{COMSOL Documentation}}} (\bibinfo {year} {2021}),\
  \bibinfo {note} {[Online; accessed 2. Jun. 2021]}\BibitemShut {NoStop}%
\bibitem [{\citenamefont {De~Witt}\ and\ \citenamefont
  {Jensen}(1953)}]{dewitt1953UeberDrehimpulsMultipolstrahlung}%
  \BibitemOpen
  \bibfield  {author} {\bibinfo {author} {\bibfnamefont {C.~M.}\ \bibnamefont
  {De~Witt}}\ and\ \bibinfo {author} {\bibfnamefont {J.~H.~D.}\ \bibnamefont
  {Jensen}},\ }\bibfield  {title} {\bibinfo {title}
  {{{\ifmmode\ddot{U}\else\"{U}\fi}ber den Drehimpuls der Multipolstrahlung}},\
  }\href {https://doi.org/10.1515/zna-1953-0409} {\bibfield  {journal}
  {\bibinfo  {journal} {Zeitschrift f{\ifmmode\ddot{u}\else\"{u}\fi}r
  Naturforschung A}\ }\textbf {\bibinfo {volume} {8}},\ \bibinfo {pages} {267}
  (\bibinfo {year} {1953})}\BibitemShut {NoStop}%
\bibitem [{\citenamefont {Li}\ \emph {et~al.}(2014{\natexlab{b}})\citenamefont
  {Li}, \citenamefont {Wu}, \citenamefont {Qu}, \citenamefont {Li},
  \citenamefont {Bai},\ and\ \citenamefont
  {Gong}}]{li2014RadiationTorqueExerted}%
  \BibitemOpen
  \bibfield  {author} {\bibinfo {author} {\bibfnamefont {Z.}~\bibnamefont
  {Li}}, \bibinfo {author} {\bibfnamefont {Z.}~\bibnamefont {Wu}}, \bibinfo
  {author} {\bibfnamefont {T.}~\bibnamefont {Qu}}, \bibinfo {author}
  {\bibfnamefont {H.}~\bibnamefont {Li}}, \bibinfo {author} {\bibfnamefont
  {L.}~\bibnamefont {Bai}},\ and\ \bibinfo {author} {\bibfnamefont
  {L.}~\bibnamefont {Gong}},\ }\bibfield  {title} {\bibinfo {title} {Radiation
  torque exerted on a uniaxial anisotropic sphere: {{Effects}} of various
  parameters},\ }\href {https://doi.org/10.1016/j.optlastec.2014.05.026}
  {\bibfield  {journal} {\bibinfo  {journal} {Optics \& Laser Technology}\
  }\textbf {\bibinfo {volume} {64}},\ \bibinfo {pages} {269} (\bibinfo {year}
  {2014}{\natexlab{b}})}\BibitemShut {NoStop}%
\bibitem [{\citenamefont {Li}\ \emph {et~al.}(2011)\citenamefont {Li},
  \citenamefont {Wu},\ and\ \citenamefont
  {Shang}}]{li2011CalculationRadiationForces}%
  \BibitemOpen
  \bibfield  {author} {\bibinfo {author} {\bibfnamefont {Z.-J.}\ \bibnamefont
  {Li}}, \bibinfo {author} {\bibfnamefont {Z.-S.}\ \bibnamefont {Wu}},\ and\
  \bibinfo {author} {\bibfnamefont {Q.-C.}\ \bibnamefont {Shang}},\ }\bibfield
  {title} {\bibinfo {title} {Calculation of radiation forces exerted on a
  uniaxial anisotropic sphere by an off-axis incident {{Gaussian}} beam},\
  }\href {https://doi.org/10.1364/OE.19.016044} {\bibfield  {journal} {\bibinfo
   {journal} {Optics Express}\ }\textbf {\bibinfo {volume} {19}},\ \bibinfo
  {pages} {16044} (\bibinfo {year} {2011})}\BibitemShut {NoStop}%
\bibitem [{\citenamefont {Li}\ \emph {et~al.}(2012{\natexlab{b}})\citenamefont
  {Li}, \citenamefont {Wu}, \citenamefont {Shang}, \citenamefont {Bai},\ and\
  \citenamefont {Cao}}]{li2012CalculationRadiationForce}%
  \BibitemOpen
  \bibfield  {author} {\bibinfo {author} {\bibfnamefont {Z.-J.}\ \bibnamefont
  {Li}}, \bibinfo {author} {\bibfnamefont {Z.-S.}\ \bibnamefont {Wu}}, \bibinfo
  {author} {\bibfnamefont {Q.-C.}\ \bibnamefont {Shang}}, \bibinfo {author}
  {\bibfnamefont {L.}~\bibnamefont {Bai}},\ and\ \bibinfo {author}
  {\bibfnamefont {C.-H.}\ \bibnamefont {Cao}},\ }\bibfield  {title} {\bibinfo
  {title} {Calculation of radiation force and torque exerted on a uniaxial
  anisotropic sphere by an incident {{Gaussian}} beam with arbitrary
  propagation and polarization directions},\ }\href
  {https://doi.org/10.1364/OE.20.016421} {\bibfield  {journal} {\bibinfo
  {journal} {Optics Express}\ }\textbf {\bibinfo {volume} {20}},\ \bibinfo
  {pages} {16421} (\bibinfo {year} {2012}{\natexlab{b}})}\BibitemShut {NoStop}%
\bibitem [{\citenamefont {Barton}\ \emph {et~al.}(1989)\citenamefont {Barton},
  \citenamefont {Alexander},\ and\ \citenamefont {Schaub}}]{Barton1989}%
  \BibitemOpen
  \bibfield  {author} {\bibinfo {author} {\bibfnamefont {J.~P.}\ \bibnamefont
  {Barton}}, \bibinfo {author} {\bibfnamefont {D.~R.}\ \bibnamefont
  {Alexander}},\ and\ \bibinfo {author} {\bibfnamefont {S.~A.}\ \bibnamefont
  {Schaub}},\ }\bibfield  {title} {\bibinfo {title} {Theoretical determination
  of net radiation force and torque for a spherical particle illuminated by a
  focused laser beam},\ }\href {https://doi.org/10.1063/1.343813} {\bibfield
  {journal} {\bibinfo  {journal} {Journal of Applied Physics}\ }\textbf
  {\bibinfo {volume} {66}},\ \bibinfo {pages} {4594} (\bibinfo {year}
  {1989})},\ \Eprint {https://arxiv.org/abs/cs/9605103} {arXiv:cs/9605103}
  \BibitemShut {NoStop}%
\bibitem [{\citenamefont {Barton}\ \emph {et~al.}(1988)\citenamefont {Barton},
  \citenamefont {Alexander},\ and\ \citenamefont {Schaub}}]{Barton1988}%
  \BibitemOpen
  \bibfield  {author} {\bibinfo {author} {\bibfnamefont {J.~P.}\ \bibnamefont
  {Barton}}, \bibinfo {author} {\bibfnamefont {D.~R.}\ \bibnamefont
  {Alexander}},\ and\ \bibinfo {author} {\bibfnamefont {S.~A.}\ \bibnamefont
  {Schaub}},\ }\bibfield  {title} {\bibinfo {title} {Internal and near-surface
  electromagnetic fields for a spherical particle irradiated by a focused laser
  beam},\ }\href {https://doi.org/10.1063/1.341811} {\bibfield  {journal}
  {\bibinfo  {journal} {Journal of Applied Physics}\ }\textbf {\bibinfo
  {volume} {64}},\ \bibinfo {pages} {1632} (\bibinfo {year}
  {1988})}\BibitemShut {NoStop}%
\bibitem [{\citenamefont {Landau}\ \emph {et~al.}(2013)\citenamefont {Landau},
  \citenamefont {Bell}, \citenamefont {Kearsley}, \citenamefont {Pitaevskii},
  \citenamefont {Lifshitz},\ and\ \citenamefont
  {Sykes}}]{landau2013electrodynamics}%
  \BibitemOpen
  \bibfield  {author} {\bibinfo {author} {\bibfnamefont {L.~D.}\ \bibnamefont
  {Landau}}, \bibinfo {author} {\bibfnamefont {J.}~\bibnamefont {Bell}},
  \bibinfo {author} {\bibfnamefont {M.}~\bibnamefont {Kearsley}}, \bibinfo
  {author} {\bibfnamefont {L.}~\bibnamefont {Pitaevskii}}, \bibinfo {author}
  {\bibfnamefont {E.}~\bibnamefont {Lifshitz}},\ and\ \bibinfo {author}
  {\bibfnamefont {J.}~\bibnamefont {Sykes}},\ }\href@noop {} {\emph {\bibinfo
  {title} {Electrodynamics of continuous media}}},\ Vol.~\bibinfo {volume} {8}\
  (\bibinfo  {publisher} {elsevier},\ \bibinfo {year} {2013})\BibitemShut
  {NoStop}%
\bibitem [{\citenamefont {toftul}(2022)}]{toftul_github_rotation}%
  \BibitemOpen
  \bibfield  {author} {\bibinfo {author} {\bibnamefont {toftul}},\ }\href
  {https://github.com/toftul/tensors-in-curvilinear-coordinates} {\bibinfo
  {title} {{tensors-in-curvilinear-coordinates}}} (\bibinfo {year} {2022}),\
  \bibinfo {note} {[Online; accessed 18. Jul. 2022]}\BibitemShut {NoStop}%
\bibitem [{\citenamefont {Bergfeld}\ and\ \citenamefont
  {Daum}(2003)}]{Bergfeld2003Jan}%
  \BibitemOpen
  \bibfield  {author} {\bibinfo {author} {\bibfnamefont {S.}~\bibnamefont
  {Bergfeld}}\ and\ \bibinfo {author} {\bibfnamefont {W.}~\bibnamefont
  {Daum}},\ }\bibfield  {title} {\bibinfo {title} {{Second-Harmonic Generation
  in GaAs: Experiment versus Theoretical Predictions of
  ${\ensuremath{\chi}}_{xyz}^{(2)}$}},\ }\href
  {https://doi.org/10.1103/PhysRevLett.90.036801} {\bibfield  {journal}
  {\bibinfo  {journal} {Phys. Rev. Lett.}\ }\textbf {\bibinfo {volume} {90}},\
  \bibinfo {pages} {036801} (\bibinfo {year} {2003})}\BibitemShut {NoStop}%
\bibitem [{\citenamefont {Koshelev}(2022)}]{koshelev2022advanced}%
  \BibitemOpen
  \bibfield  {author} {\bibinfo {author} {\bibfnamefont {K.}~\bibnamefont
  {Koshelev}},\ }\emph {\bibinfo {title} {Advanced trapping of light in
  resonant dielectric metastructures for nonlinear optics}},\ \href@noop {}
  {Ph.D. thesis},\ \bibinfo  {school} {Australian National University}
  (\bibinfo {year} {2022})\BibitemShut {NoStop}%
\bibitem [{\citenamefont {Kristensson}(2014)}]{kristensson2014spherical}%
  \BibitemOpen
  \bibfield  {author} {\bibinfo {author} {\bibfnamefont {G.}~\bibnamefont
  {Kristensson}},\ }\href@noop {} {\bibinfo {title} {Spherical vector waves}}
  (\bibinfo {year} {2014})\BibitemShut {NoStop}%
\bibitem [{\citenamefont {Weiss}\ and\ \citenamefont
  {Muljarov}(2018)}]{Weiss2018Aug}%
  \BibitemOpen
  \bibfield  {author} {\bibinfo {author} {\bibfnamefont {T.}~\bibnamefont
  {Weiss}}\ and\ \bibinfo {author} {\bibfnamefont {E.~A.}\ \bibnamefont
  {Muljarov}},\ }\bibfield  {title} {\bibinfo {title} {{How to calculate the
  pole expansion of the optical scattering matrix from the resonant states}},\
  }\href {https://doi.org/10.1103/PhysRevB.98.085433} {\bibfield  {journal}
  {\bibinfo  {journal} {Phys. Rev. B}\ }\textbf {\bibinfo {volume} {98}},\
  \bibinfo {pages} {085433} (\bibinfo {year} {2018})}\BibitemShut {NoStop}%
\bibitem [{\citenamefont {Muljarov}\ and\ \citenamefont
  {Weiss}(2018)}]{Muljarov2018May}%
  \BibitemOpen
  \bibfield  {author} {\bibinfo {author} {\bibfnamefont {E.~A.}\ \bibnamefont
  {Muljarov}}\ and\ \bibinfo {author} {\bibfnamefont {T.}~\bibnamefont
  {Weiss}},\ }\bibfield  {title} {\bibinfo {title} {{Resonant-state expansion
  for open optical systems: generalization to magnetic, chiral, and
  bi-anisotropic materials}},\ }\href {https://doi.org/10.1364/OL.43.001978}
  {\bibfield  {journal} {\bibinfo  {journal} {Opt. Lett.}\ }\textbf {\bibinfo
  {volume} {43}},\ \bibinfo {pages} {1978} (\bibinfo {year}
  {2018})}\BibitemShut {NoStop}%
\bibitem [{\citenamefont {Perrin}(1934)}]{Perrin1934Oct}%
  \BibitemOpen
  \bibfield  {author} {\bibinfo {author} {\bibfnamefont {F.}~\bibnamefont
  {Perrin}},\ }\bibfield  {title} {\bibinfo {title} {{Mouvement brownien d'un
  ellipsoide - I. Dispersion di{\ifmmode\acute{e}\else\'{e}\fi}lectrique pour
  des mol{\ifmmode\acute{e}\else\'{e}\fi}cules ellipsoidales}},\ }\href
  {https://doi.org/10.1051/jphysrad:01934005010049700} {\bibfield  {journal}
  {\bibinfo  {journal} {J. Phys. Radium}\ }\textbf {\bibinfo {volume} {5}},\
  \bibinfo {pages} {497} (\bibinfo {year} {1934})}\BibitemShut {NoStop}%
\bibitem [{\citenamefont {Cantor}\ and\ \citenamefont
  {Schimmel}(1980)}]{cantor1980biophysical}%
  \BibitemOpen
  \bibfield  {author} {\bibinfo {author} {\bibfnamefont {C.~R.}\ \bibnamefont
  {Cantor}}\ and\ \bibinfo {author} {\bibfnamefont {P.~R.}\ \bibnamefont
  {Schimmel}},\ }\href@noop {} {\emph {\bibinfo {title} {Biophysical chemistry:
  Part II: Techniques for the study of biological structure and function}}}\
  (\bibinfo  {publisher} {Macmillan},\ \bibinfo {year} {1980})\BibitemShut
  {NoStop}%
\bibitem [{\citenamefont {Koenig}(1975)}]{koenig1975brownian}%
  \BibitemOpen
  \bibfield  {author} {\bibinfo {author} {\bibfnamefont {S.~H.}\ \bibnamefont
  {Koenig}},\ }\bibfield  {title} {\bibinfo {title} {Brownian motion of an
  ellipsoid. a correction to perrin's results},\ }\href@noop {} {\bibfield
  {journal} {\bibinfo  {journal} {Biopolymers: Original Research on
  Biomolecules}\ }\textbf {\bibinfo {volume} {14}},\ \bibinfo {pages} {2421}
  (\bibinfo {year} {1975})}\BibitemShut {NoStop}%
\bibitem [{\citenamefont {Landau}\ and\ \citenamefont
  {Lifshitz}(2013)}]{landau2013fluid}%
  \BibitemOpen
  \bibfield  {author} {\bibinfo {author} {\bibfnamefont {L.~D.}\ \bibnamefont
  {Landau}}\ and\ \bibinfo {author} {\bibfnamefont {E.~M.}\ \bibnamefont
  {Lifshitz}},\ }\href@noop {} {\emph {\bibinfo {title} {Fluid Mechanics:
  Landau and Lifshitz: Course of Theoretical Physics, Volume 6}}},\
  Vol.~\bibinfo {volume} {6}\ (\bibinfo  {publisher} {Elsevier},\ \bibinfo
  {year} {2013})\BibitemShut {NoStop}%
\bibitem [{\citenamefont {Coutsias}\ and\ \citenamefont
  {Romero}(2004)}]{Coutsias2004}%
  \BibitemOpen
  \bibfield  {author} {\bibinfo {author} {\bibfnamefont {E.~A.}\ \bibnamefont
  {Coutsias}}\ and\ \bibinfo {author} {\bibfnamefont {L.}~\bibnamefont
  {Romero}},\ }\bibfield  {title} {\bibinfo {title} {{The Quaternions with an
  application to Rigid Body Dynamics}},\ }\href
  {https://digitalrepository.unm.edu/math_fsp/4} {\bibfield  {journal}
  {\bibinfo  {journal} {UNM Digital Repository}\ } (\bibinfo {year}
  {2004})}\BibitemShut {NoStop}%
\end{thebibliography}%

\end{document}